# Skin Cancer Segmentation and Classification Using Vision Transformer for Automatic Analysis in Dermatoscopy-based Non-invasive Digital System

**Galib Muhammad Shahriar Himel** [1(Corresponding author)], **ORCid:** **0000-0002-2257-6751**
**Md. Masudul Islam** [1],
**Kh Abdullah Al-Aff** [2],
**Shams Ibne Karim** [3],
**Md. Kabir Uddin Sikder** [1]
[1]Jahangirnagar University, Dhaka, Bangladesh
[2]Bangladesh University of Health Sciences, Dhaka, Bangladesh
[3]Bangabandhu Sheikh Mujib Medical University, Dhaka, Bangladesh
[*]**Corresponding Author:** galib.muhammad.shahriar@gmail.com

**Abstract:**
Skin cancer is a significant health concern worldwide, and early and accurate diagnosis plays a crucial role in improving patient outcomes. In recent years, deep learning models have shown remarkable success in various computer vision tasks, including image classification. In this research study, we introduce an approach for skin cancer classification using Vision Transformer, a state-of-the-art deep learning architecture that has demonstrated exceptional performance in diverse image analysis tasks. The study utilizes the HAM10000 dataset; a publicly available dataset comprising 10,015 skin lesion images classified into two categories: benign (6705 images) and malignant (3310 images). This dataset consists of high-resolution images captured using Dermatoscopes and carefully annotated by expert dermatologists. Preprocessing techniques, such as normalization and augmentation, are applied to enhance the robustness and generalization of the model. The Vision Transformer architecture is adapted to the skin cancer classification task. The model leverages the self-attention mechanism to capture intricate spatial dependencies and long-range dependencies within the images, enabling it to effectively learn relevant features for accurate classification. Segment Anything Model (SAM) is employed to segment the cancerous areas from the images; achieving an IOU of 96.01% and Dice Coefficient of 98.14% and then various pre-trained models are used for classification using Vision Transformer architecture. Extensive experiments and evaluations are conducted to assess the performance of our approach. The results demonstrate the superiority of the Vision Transformer model over traditional deep learning architectures in skin cancer classification in general with some exceptions. Upon experimenting on six different models: ViT-Google, ViT-MAE, ViT-ResNet50, ViT-VAN, ViT-BEiT, and ViT-DiT we found out that the ML approach achieves 96.15% accuracy using Google's ViT patch-32 model with a low false negative ratio on the test dataset; showcasing its potential as an effective tool for aiding dermatologists in the diagnosis of skin cancer.

## 1. Introduction

Cancer is a condition that arises when cells undergo uncontrolled division and extend into nearby tissues. The development of cancer is triggered by alterations and mutations in the DNA. The majority of DNA changes responsible for cancer occur within specific regions known as genes. Among the various types of cancers, skin cancer is among the five on the list. If we disregard breast and prostate cancer which are gender-dependent, skin cancer will remain in the third largest cancer category among many others. Based on the statistics released by the American Cancer Society (ACS) [1], there were 58,120 recorded cases of skin cancer among males and 39,490 cases among females. An intriguing observation is that the incidence of skin cancer has been steadily rising from 1992 to 2019, with a notable exception in 2020 [2]. This exception can be attributed to the understandable decrease in cases during the COVID-19 pandemic, as people were mostly confined to their homes. This decline is reasonable considering that exposure to ultraviolet (UV) radiation is a significant contributing factor to the development of skin cancer.
More people are diagnosed with skin cancer each year in the U.S. than all other cancers combined [3]. More than 5,400 people worldwide die of nonmelanoma skin cancer every month [4]. The number of melanoma deaths is expected to increase by 4.4 percent in 2023 [3]. More people develop skin cancer because of indoor tanning than develop lung cancer because of smoking [5]. Skin cancer represents approximately 2 to 4 percent of all cancers in Asians, 4 to 5 percent of all cancers in Hispanics, and 1 to 2 percent of all cancers in Black people [6][7][8]. Skin cancers account for 3 percent of pediatric cancers [9].

There are various factors involving the initiation of developing skin cancer such as Ultra Violet rays, arsenic consumption, cigarette smoking, exposure to nuclear radiation, excessive X-ray exposure, indoor tanning, drinking alcohol, viruses, changing environments, sunburn, abnormal swelling, eczema, weak immune system, Human Papilloma Virus, etc. There are also environmental factors that work as a catalyst for developing skin cancer such as working coal, tar, petroleum, and shale oils, etc. [10]. Some diseases may fasten the development rate of skin cancer but not necessarily the root cause of cancer themselves such as AIDS and other diseases that involve the weakening of the immune system.

Skin cancer may form on the upper part of the skin (which is visible to us) as well as in the inner parts of different layers of the skin. More than 90% of skin cancers are caused by direct sunlight exposure, to be specific UV exposure. Almost all of the cases are related to cancer development on the upper visible part of the skin. The affected skin areas can be easily captured by a dermatoscope or high-resolution smartphone camera. On the other hand, skin cancers that develop inside the skin layers require invasive sample collection methods. Since more than 90% of skin cancer is caused by UV rays and these cancers occur primarily on the surface of the skin through UV exposure, it can be said that the majority of skin cancers can be identified through non-invasive methods using a Dermatoscope, as it allows for the collection of samples from the skin.

Traditionally, the ABCDE [11] method is used for the primary diagnosis (classification) of skin cancer. The ABCDE acronym is used as a mnemonic for recognizing potential signs of melanoma, a type of skin cancer that is shown in **Figure 1**. Each letter corresponds to a characteristic feature that may indicate the presence of melanoma. Here is the breakdown of the ABCDE criteria:

- A - Asymmetry: Melanomas often exhibit irregular or asymmetrical shapes, where one half does not match the other half.
- B - Border irregularity: The borders of a melanoma may be uneven, ragged, or notched, rather than smooth and well-defined.
- C - Color variation: Melanomas can have a range of colors within the same lesion, such as different shades of brown, black, blue, red, or white.
- D - Diameter: While melanomas can be smaller, any mole or lesion with a diameter larger than 6 millimeters (about the size of a pencil eraser) should be closely examined.
- E - Evolution or changes over time: Watch out for any changes in size, shape, color, elevation, or other characteristics of a mole or lesion.

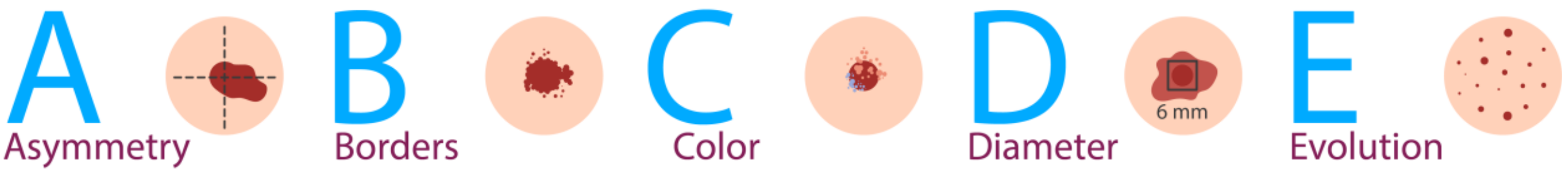

**Figure 1.** ABCDE method for primary diagnosis of skin cancer

After the primary classification, the sample is transferred to the pathological analysis team for a definitive result. The pathological analysis is an invasive method. This method can be used for classifying all kinds of cancer cells. However, it has its disadvantages. This method is very costly, time-consuming, and painful. Pathological diagnosis of skin cancer typically involves an invasive method known as a biopsy. The biopsy procedure is performed by a healthcare professional, usually a dermatologist or a surgeon, and it involves the removal of a small sample of suspicious skin tissue for further examination.

There are different types of skin biopsies, including:

- Excisional Biopsy: This type of biopsy involves the complete removal of the suspicious skin lesion, along with a small margin of surrounding healthy tissue. The excised sample is then sent to a pathology laboratory for analysis.
- Incisional Biopsy: In this method, only a portion of the suspicious skin lesion is removed for examination. It is typically done when the lesion is large or deep, and the entire lesion cannot be easily excised.
- Punch Biopsy: A punch biopsy involves using a special tool called a punch to remove a small circular piece of skin tissue, including the suspicious lesion and a small portion of normal skin around it.
- Shave Biopsy: This method involves shaving off the top layers of the skin using a surgical blade to obtain a sample. It is commonly used for superficial lesions or those located on the surface of the skin.

Once the skin sample is obtained through the biopsy procedure, it is sent to a pathology laboratory, where it undergoes microscopic examination by a pathologist. The pathologist examines the tissue sample, looking for characteristic features of skin cancer, such as abnormal cell growth patterns, cellular atypia, and invasion of surrounding tissues. Based on the examination findings, the pathologist provides a definitive diagnosis and may further classify the specific type and stage of skin cancer.
Due to the time-consuming nature of even the primary diagnosis process used in the conventional method, it often becomes very difficult to identify cancer within the target time. At the same time handling a lot of patients within the expected timeframe becomes a nightmare. From the statistics, it is also clear that day by day skin cancer patients are increasing rapidly as a result of the depression of the ozone layer creating an inability to block the UV rays. So, soon the hospital and the diagnosis centers will face more difficulties in managing the situation in due time. To solve this issue artificial intelligence can play a vital role. Using a Dermatoscope the suspected areas of the skin can be captured at up to 20X magnification very easily. These images can be used to train an intelligent system that can classify cancer cells accurately. Computers, smartphones, or any other smart devices can be used to do the task.

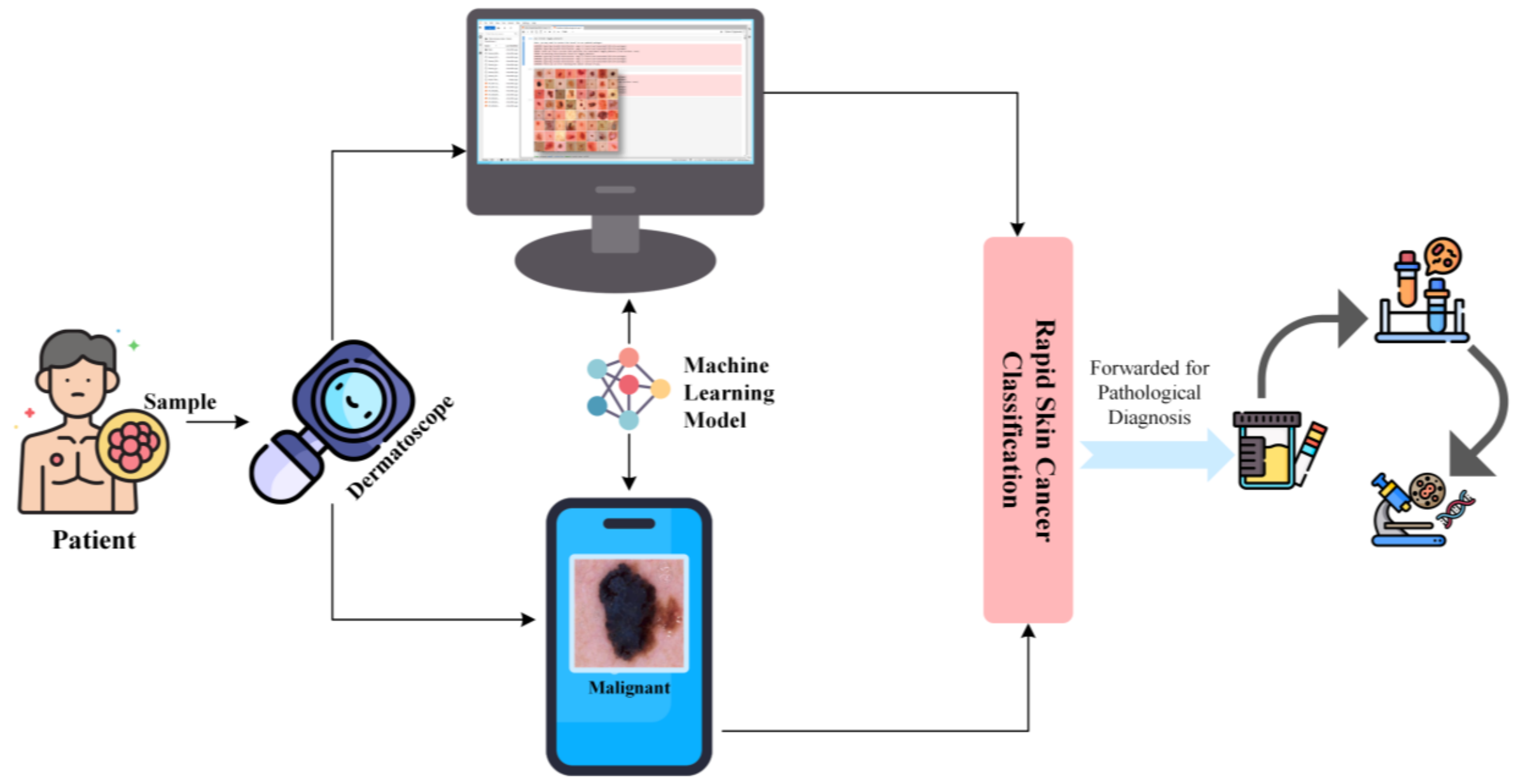

**Figure 2.** Rapid Cancer Cell Classification Using Dermatoscope

In this paper, we are proposing a vision transfer-based machine learning model that will classify the cancerous cells on the upper part of the skin. This model can be mounted on any smart device which can be used to capture images using a Dermatoscope and instantly generate a classification decision regarding the suspected area. In **Figure 2**, the workflow of rapid cancer cell classification is shown.
In this research, the HAM10000 [12] dataset is used for training the model. At first, we manually annotated the suspected areas and then applied the Segment Anything Model (SAM) [13] to train the segmentation model. This model will be used to detect the specific area using the concept of semantic segmentation. In the second stage, the classification model will be used to differentiate between the benign and malignant categories. The major contribution of our research:

- Creating the segmentation model using the Segment Anything Model (SAM)
- For classification, we have used Google's ViT model (patch 32) [14], Visual Attention Network (VAN) [15], ResNet50 [14], Facebook's MAE model [16], Document Image Transformer (DiT) [17], BERT pre-training of Image Transformer (BEiT) [18] to train the Vision Transformer model. Comparing these models, we have found out that Google's ViT model (patch-32) generates better results than the other models in skin cancer classification.
- We meticulously fine-tuned the hyperparameters for our model, exploring various optimization strategies, learning rates, and loss functions. Specifically, we evaluated the performance under different optimizers, including 'Adam', 'Adamax', 'RMSProp', and 'SGD'. The learning rates considered spanned a range of values: 0.1, 0.01, 0.001, 0.0001, 0.00002, and 0.00001. Additionally, we experimented with different loss functions, namely 'Cross Entropy Loss' and 'Hinge'. After a thorough analysis, we determined that the most effective combination for our Vision Transformer (ViT) model involved employing the 'Adam' optimizer with a learning rate of 0.00002 and using 'Cross Entropy Loss' for calculating the loss. These optimized hyperparameters were instrumental in achieving superior performance and model efficiency.

The following paper is structured in the following way: Section **2** describes the existing literature on the topic and comparison among them. Section **3** describes the methodology of our works which includes dataset description, experimental model, and experimental setup. Section **4** provides various graphical representations of the result analysis and a discussion of the challenging part of our study. Section **5** provides the concluding remarks.

## 2. Related Works

Traditionally, the initial step of the skin cancer classification process involves the patient visiting a dermatologist or a healthcare professional with concerns about a suspicious skin lesion. The specialist conducts a thorough examination of the skin, including a visual inspection of the lesion and an examination of the patient's medical history. If the specialist suspects a potential skin cancer, they may use a dermoscopy, a handheld device with magnification and lighting, to obtain a closer view of the lesion. Dermoscopy allows for a detailed examination of the lesion's surface structures, patterns, and colors, aiding in the assessment of malignancy. This takes around 10-15 minutes. To obtain a definitive diagnosis, a skin biopsy is often performed. The specialist numbs the area surrounding the lesion with a local anesthetic and extracts a sample of skin tissue, including a portion of the suspicious lesion. Various biopsy techniques may be used, such as punch biopsy, incisional biopsy, or excisional biopsy, depending on the size and characteristics of the lesion. Usually, it takes 2-3 weeks. The skin tissue sample obtained from the biopsy is sent to a pathology laboratory for histopathological examination. A pathologist, specializing in dermatopathology, analyzes the tissue under a microscope. They assess cellular characteristics, such as cell shape, size, and organization, to determine if the lesion is cancerous or benign. This process takes 2-3 days. The examination also helps in identifying the specific type of skin cancer, such as melanoma, basal cell carcinoma, squamous cell carcinoma, or other variants. If skin cancer is confirmed, the pathologist and specialist determine the stage and grade of the cancer based on additional tests and evaluations. Staging involves assessing the tumor size, depth of invasion, involvement of lymph nodes, and potential spread to distant sites. Grading refers to the assessment of tumor aggressiveness and differentiation.

To know the decision definitively, it takes almost a month which may be hazardous if the patient is in the final stage. To minimize the time taken for diagnosis research has been done in this field for decades. Artificial Intelligence has been used in this field in recent years. Since artificial intelligence is a new field, achieving highly accurate results in skin cancer diagnosis has not been possible in the early stages of the research. Continuous development in machine learning algorithms has been done to gain better accuracy. Over the years, pathologists, doctors, and computer scientists collaborated along sides to develop a better digital classification system. We have presented some prominent research conducted on skin cancer detection from 2018 to 2023.

Dorj et al. [21] utilized a dataset from the internet comprising 3753 images with four classes. Employing AlexNet and ECOC SVM, they achieved a notable accuracy of 94.2%. AlexNet was employed for feature extraction, while ECOC SVM handled the classification task. It's essential to note that the dataset collected from the internet didn't adhere to benchmark standards. In a different approach, Rezvantalab et al. [22] utilized the HAM10000 (10015 images) + PH2 (120 images) dataset, consisting of eight classes. Their model, DenseNet 201, achieved an accuracy of 86.59%. The authors incorporated various pre-trained models, including DenseNet 201, ResNet 152, Inception v3, and InceptionResNetV2. Results were presented through AUC values ranging from 93.80% to 99.3% across eight categories. Moving to the PH2 dataset with three classes, K. M. Hosny et al. [23] achieved an accuracy of 98.61% using a Customized AlexNet. The augmentation of the main dataset resulted in 4400 images, and a modified version of AlexNet was applied to obtain the results. Dascalu et al. [24] delved into the ISIC 2017 dataset (5161 images) with two classes, obtaining an AUC of 81.40%. Their unique approach involved using K-Means Clustering and sonification to evaluate the impact of image quality on diagnosis accuracy. In another study, T. C. Pham et al. [25] explored the HAM10000 (1113 images) and ISIC 2016 (172 images) datasets, comprising one class. Their approach, involving Linear Normalization, HSV, and LBP Balanced Random Forest, achieved an accuracy of 74.75%. This study was a comparative analysis of the color, texture, and shape features of melanoma skin cancer cells. Hekler et al. [26], dealing with the HAM10000 & ISIC dataset (11,444 images combined) with five classes, reached an accuracy of 82.95% through a fusion of Physician and CNN. Notably, the study presented results for both multiclass and binary classifications, utilizing the XGBoost algorithm for the former. T. Emara et al. [27] tackled the HAM10000 dataset (7 classes) and achieved an accuracy of 94.7% using a Modified InceptionV4 model. Their study focused on proposing a modified InceptionV4 model tailored to the unbalanced proportions of the HAM10000 dataset. Chaturvedi et al. [28] applied a MobileNet pre-trained model on the HAM10000 dataset (7 classes) and achieved an accuracy of 83.1%. The study involved training on an augmented dataset comprising 38,569 images. In a similar vein, Mohapatra et al. [29] utilized a MobileNet pre-trained model on the HAM10000 dataset (7 classes) without modifications and achieved an accuracy of 80%. Moving to the N/A dataset with nine classes, Chen et al. [30] achieved an accuracy of 83.74% using a ResNet50 pre-trained model. The study emphasized the application of the ResNet50 model to achieve accuracy across nine classes of skin lesions. Jinnai et al. [31] worked with the National Cancer Center, Tokyo dataset (5846 images) with six classes. Their approach involved FRCNN, BCD, and TRN, resulting in accuracies of 86.2%, 79.5%, and 75.1%, respectively. The study utilized a customized dataset with two main classes (benign and malignant) and compared results across different classifiers. Chaturvedi et al. [32] explored the HAM10000 dataset (7 classes) and achieved an accuracy of 92.83% using ResNetXt101. The study extensively researched optimal hyper-parameter configurations for five pre-trained models on ImageNet, with ResNetXt101 yielding the best performance. Garg et al. [33] worked with the HAM10000 dataset (7 classes), achieving an accuracy of 90.51% using ResNet50. Their study applied two pre-trained models, VGG16 and ResNet50, along with three meta-learners: Random Forest, XGBoost, and SVM. Benedetti et al. [34] applied the HAM10000 dataset (7 classes) and achieved an accuracy of 78.9% using a Modified InceptionResNetv2. Their model incorporated an InceptionResNetv2 architecture with the addition of a flattening layer. Moving to the ISIC 2019 dataset (25331 images) with nine classes, N. Gouda & Amudha [35] achieved an accuracy of 92% using ResNet34. Their study focused on applying the ResNet34 model for the classification of cancerous cells. Ismail et al. [36] worked with the HAM10000 dataset (7 classes) and achieved an accuracy of 84.01% using a combination of ResNet50, VGG16, and DenseNet. The authors ran ResNet50, VGG16, and DenseNet in parallel, concatenating their results by stacking. Kondaveeti & and Edupuganti [37] proposed a model with ResNet50, MobileNet, Xception, and InceptionV3 as base models for the classification of skin lesion images in the HAM10000 dataset. They achieved an accuracy of 90%. Maiti et al. [38] applied the GAN Data dataset (4992 images) with three classes and achieved an accuracy of 97.08% using Random Forest. Their study suggested that employing a color quantization technique combined with synthetic data generation significantly enhanced the accuracy of well-known machine learning models. R. Ashraf et al. [39] utilized the DHQ Hospital dataset (400 images) with three classes and achieved an accuracy of 93.29% using DCNN. The study emphasized the importance of skin cancer segmentation and image preprocessing. Pacheco & Krohling [40] worked with the ISIC 2019 dataset (33,569 images) with eight classes, achieving an accuracy of 91.3%. They compared the MetaBlock approach with models without using metadata, the baseline concatenation method, and the MetaNet. Alagu & Bagan [41] utilized the ISIC dataset (500 images) with three classes and achieved an accuracy of 95% using CNN and DenseNet. The study focused on identifying melanoma cells, and the augmented data were trained using DenseNet. Krohling et al. [42] applied the PAD-UFES-20 dataset (2057 images) with seven classes and achieved an accuracy of 85% using ResNet50 and the Differential Evolution (DE) algorithm. The study highlighted the importance of data balancing for success. Mijwil [43] worked with the ISIC dataset (24000+ images) with two classes, achieving an accuracy of 86.90% using InceptionV3. The authors used InceptionV3, ResNet, and VGG19 to apply CNN to the dataset and reported the best result. M. Shah [44] applied the HAM10000 dataset 7 classes and achieved an accuracy of 90.6% using LRNet. The study employed LRNet to develop the ML model for classifying skin cancers. Maron et al. [45] explored the SAM dataset (319 images), SAM-C, and SAM-P dataset with two classes. They applied AlexNet, VGG16+BN, ResNet50, and DenseNet121, although specific accuracy values were not detailed. The study focused on assessing the robustness of these AI methods by comparing their application to an Unmodified dataset (SAM) and artificially modified datasets (SAM-C, SAM-P). Ali et al. [46] utilized the HAM10000 dataset (2 classes) and achieved an accuracy of 91.93% using a modified DCNN. The authors compared their results with AlexNet, ResNet, VGG-16, DenseNet, and MobileNet. Dascalu et al. [47] worked with a dataset comprising images from HAM10000 (149 images), Dascalu et al. (16 images) [24], Biopsy-validated (159 images), and JID2018 (39 images) with two classes. They achieved an accuracy of DI-87.8% using CNN and Sonification. The study involved a model where a CNN predicted malignancy from a raw image overlaid with a second independent CNN processing sonification of the original image, combined into a unified malignancy classifier. Yilmaz et al. [48] applied the ISIC 2017 dataset (2750 images) with two classes and achieved an accuracy of 82% using NASNetMobile. The study utilized MobileNet, MobileNetV2, and NASNetMobile in different batch sizes to optimize results. Ahmad et al. [49] worked with the HAM10000 dataset (4 classes) and achieved an accuracy of 92.5% using TED-GAN and CNN. The study proposed a framework called T-Distribution Encoder & Decoder-Generative Adversarial Network to detect melanoma skin cancer. Kausar et al. [50] utilized the ISIC 2019 dataset (25331 images) with eight classes and achieved an accuracy of 98.6% using a Weighted Average and voting Ensemble of five pre-trained models. The researchers employed ResNet, InceptionV3, DenseNet, InceptionResNetV2, and VGG-19 models, combining their predictions using majority voting and weighted majority voting for enhanced accuracy. Mazur, B. et al. [51] employed the ISIC 2018 Dataset, consisting of 33,900 images with two classes, achieving an accuracy of 83.4%. Their methodology involved utilizing Six Pre-trained Models, Grad-CAM, and the UMAP algorithm. The authors introduced the DUNEScan (Deep Uncertainty Estimation for Skin Cancer) web server, which conducts a comprehensive analysis of uncertainty in prevalent skin cancer classification models based on convolutional neural networks (CNNs). The server incorporates six efficient CNN models, including the winners of the dermatological Kaggle competition: Inceptionv313, ResNet5014, MobileNetv23, EfficientNet15, BYOL16, and SwAV17. Bechelli and Delhommelle [52] employed the ISIC dataset comprising 3,297 images with two classes, achieving a 73% accuracy using an ensemble of LR, LDA, KNN, CART, and GNB. Their study extensively utilized Machine Learning Algorithms (LR, LDA, KNN, CART, GNB) and explored various combinations on the ISIC dataset. In contrast, Deep Learning models (Xception, VGG16, ResNet50) were employed on the HAM10000 dataset, with ResNet50 proving superior at 88% accuracy. Additionally, they achieved an 88% accuracy using a Fine-tuned VGG16 model. Maniraj & Maran [53] utilized the PH2 dataset, which comprises 200 images with three classes, achieving an impressive accuracy of 99.33% through Hybrid Deep Learning (VGG) and a Sub-band fusion of 3D wavelets. The study introduced a novel hybrid deep learning (HDL) methodology (VGG model) that incorporates 3D wavelet fusion through sub-band processing, demonstrating improved classification outcomes. Y. Filali et al. [54] employed the combined PH2 (200 images) and ISIC 2017 (2,000 images) dataset with unspecified classes, achieving an accuracy of 82% using ResNet18. The authors explored ResNet, VGG16, GoogleNet, and AlexNet for skin cancer classification. Hassan Bedeir et al. [55] used the HAM10000 dataset with seven classes, achieving an accuracy of 94.14% through a Merged ResNet50 and VGG16 approach. The study aimed at achieving high accuracy in classifying various skin cancer types using three approaches: ResNet-50, VGG-16, and a merged model combining both techniques through the concatenate function. Gouda, W. et al. [56] utilized the ISIC 2018 Dataset with 3,533 images and two classes, achieving an 85.8% accuracy using Fine-tuned CNN, ResNet, InceptionV3, and InceptionResNet. Their study employed a CNN model for identifying two main tumor categories: malignant and benign. Image enhancement using ESRGAN and fine-tuning with transfer learning models like ResNet50, InceptionV3, and Inception ResNet contributed to the achieved accuracy. Fraiwan and Faouri [57] used the HAM10000 dataset with seven classes, achieving an accuracy of 82.9% through DenseNet201. The study explored multiple models, including SqueezeNet, GoogLeNet, Inceptionv3, DenseNet-201, MobileNetv2, ResNet18, RestNet50, ResNet101, Xception, InceptionResNet, ShuffleNet, DarkNet-53, and EfficientNet-B0, to obtain the reported accuracy. Tabrizchi, H. et al. [58] applied the ISIC dataset with 33,126 images and two classes, achieving an accuracy of 86.30% using an enhanced VGG16 model. The research introduced an improved VGG16 model for the early identification of skin cancer using dermoscopic images. Naeem, A. et al. [59] used the ISIC 2019 dataset with 25,331 images and four classes, achieving an impressive accuracy of 96.91% using SCDNet. The proposed SCDNet model combined VGG16 with convolutional neural networks (CNN) for the classification of various forms of skin cancer, outperforming four widely used pre-trained classifiers in the medical field: ResNet50, InceptionV3, AlexNet, and VGG19. Huynh, A.T. et al. [60] employed the combined SIIM-ISIC 2020 + ISIC 2019 dataset with 58,457 images and two classes, achieving an AUC of 98.78% using InceptionResNetV2. The researchers combined two datasets and applied the backbones of models including EffecientNetB6, VGG16, ResNet152V2, InceptionResNetV2, and InceptionV3, evaluating performance based on Loss, AUC, and Sensitivity. Xin, C. et al. [61] utilized the HAM10000 dataset with seven classes, achieving an accuracy of 94.3% using ViT and SkinTrans. The authors proposed the SkinTrans model, an improved transformer network derived from the Visual Transformer (ViT) model, and obtained satisfactory results in two different datasets. Bassel, A. et. al. [62] used the ISIC dataset with 1,000 images and two classes, achieving a 90.9% accuracy using Xception. The classification method involved stacking classifiers using a three-fold approach, employing feature extraction with Resnet50, Xception, and VGG16. Ali, K. et al. [63] used the HAM10000 dataset with seven classes, achieving an 87.9% accuracy using EfficientNetB4. The study involved training on EfficientNet models ranging from EfficientNet B0 to B7, with EfficientNet B4 exhibiting the highest accuracy. Qasim Gilani et al. [64] employed the ISIC 2019 dataset with 6,993 images and two classes, achieving an 89.57% accuracy using spiking VGG-13. The spiking VGG-13 model utilized the surrogate gradient descent technique for classifying melanoma and non-melanoma cancer. Durães & Véstias [65] used the HAM10000 dataset with seven classes, achieving an 87% accuracy using a dual Model: ResNet18 + ResNet50. The research focused on developing a low-cost skin cancer classification system implemented on an FPGA, optimizing models with the Vitis-AI design flow and emphasizing speed rather than accuracy. Rezk, E. et al. [66] employed the PH2 dataset with 200 images and four classes, achieving an 87% accuracy using an incremental DNN and Modified InceptionV3. The researchers developed a progressive multi-output model predicting lesion source, malignancy classification, and disease diagnosis. Tembhurne, J.V. et al. [67] utilized the ISIC dataset with 2,637 images and two classes, achieving a 93% accuracy using a Voting Ensemble of LR, SVM, and Modified VGG19. The study employed an ensemble of ML and DL techniques, utilizing modified VGG19 for feature extraction and component analysis. Shaaban, S. et al. [68] used the HAM10000 dataset with two classes, achieving a 96.66% accuracy using Xception. Two optimization algorithms were employed to optimize parameters for multiple models, resulting in the highest accuracy. Tahir, M. et al. [69] employed the ISIC 2020 + HAM100000 + DermIS dataset with four classes, achieving a 94.17% accuracy using DSCC-Net. The authors introduced the DSC-Net model to address dataset imbalance and evaluated its performance against six baseline deep networks. N. S. Karpagam et al. [70] utilized the DERMIS dataset with 402 images and two classes, achieving a 97% accuracy using SVM and KNN. The SVM method was applied to the combined feature of LBO and GLCM. K. Mridha et al. [71] used the HAM10000 dataset with seven classes, achieving an 82% accuracy using an Optimized CNN. The research introduced a refined Convolutional Neural Network (CNN) with an XAI-based system incorporating Grad-CAM and Grad-CAM++. H. L. Gururaj et al. [72] employed the HAM10000 dataset with seven classes, achieving a 91.2% accuracy using DensNet169. The authors applied DensNet169 and ResNet50 models, measuring accuracy in both under-sampled and over-sampled instances and testing accuracy in different training-testing splits. They achieved an 83% accuracy using ResNet50. Khan et al. [73] introduced an HSIFT descriptor crafted for hands to extract features for anatomical object classification, achieving significant performance improvement over traditional CNN-based models. Gulzar and Khan [74] conducted a comparative study on U-Net and attention-based methods for skin lesion image segmentation, attaining 92.11% accuracy with a superior hybrid TransUNet. Khan et al. [75] proposed an ensemble (XG-Ada-RF) on Extreme Gradient Boosting, Ada-Boost, and Random Forest, achieving 95.9% accuracy for tumor detection and 94.9% for normal brain tumor images. Mehmood et al. [76] presented SBXception, a modified model for the HAM10000 dataset, achieving 96.97% accuracy on a holdout test set.
Siddique et. al. [77] researched the existing literature about U-Net on medical image segmentation. Various sectors including skin cancer diagnosis have been thoroughly reviewed in this article. Krithika et. al. [78] also conducted a systematic review of the modified architecture of U-Nets, demonstrating the applicability and limitation of U-Nets. Sreelatha et al. [79] presented a segmentation method utilizing the Gradient and Feature Adaptive Contour (GFAC) model on the PH2 dataset, achieving an accuracy of 98.64%. Liu et al. [80] applied an enhanced U-Net model with and without an ensemble for skin lesion segmentation on the ISIC 2017 dataset, reaching 92.6% accuracy without the test ensemble and 93% accuracy with the test ensemble. Wu et al. [81] introduced the C-UNet model, incorporating Inception-like convolutional blocks, recurrent convolutional blocks, and dilated convolutional layers. Their modified model achieved a Jaccard Index of 77.5% and a Dice Coefficient of 86.9% on the ISIC 2018 dataset. Tang et al. [82] utilized a Separable-UNet for skin lesion segmentation across three datasets: ISIC 2016, ISIC 2017, and PH2, achieving an average Dice coefficient of 93.03% and a Jaccard index of 89.25% for ISIC 2016, 86.93% and 79.26% for ISIC 2017, and 94.13% and 89.40% for PH2. Araújo et al. [83] combined U-Net and LinkNet methods and applied them to three datasets: PH2, ISIC 2018, and DermIS, obtaining an average Dice of 0.923 in PH2, Dice = 0.893 in ISIC 2018, and Dice = 0.879 in DermIS. Nawaz et al. [84] proposed an improved DenseNet77-based UNET model for the ISIC 2017 and ISIC 2018 datasets, achieving segmentation accuracies of 99.21% and 99.51%, respectively. In 2018, Facebook introduced Detectron2, an object detection model that has demonstrated remarkable results in recent medical segmentation research, surpassing traditional U-Net models [85]. Despite the existence of an even more advanced model from Facebook called “SAM: Segment Anything Model” it has yet to be employed in medical research. Given the success of the previous model, we have opted to utilize SAM in our research during the segmentation phase.
From the existing literature, we have come to an understanding that various pre-trained models and traditional machine learning algorithms were being used on different skin cancer datasets with or without fine-tuning and modification. Very few experiments were done using ViT while scopes of ViT are more than the traditional methods in terms of time complexity and accuracy. For complete identification of cancerous regions, the whole process is done in two steps: segmentation and classification. In medical image segmentation, various models like UNet-based models have been used throughout the years. However, rather than using conventional segmentation methods we have decided to use SAM: An object detection model developed by Facebook [13], to detect regions of interest first as the model has never been used in the field of cancer detection. After segmenting the cancerous regions, the classification can be done to identify benign and malignant cases. In this article, we have aimed to experiment on several ViT models to identify the best one to classify skin cancers from the HAM10000 dataset.

## 3. Methodology and Implementation

### *3.1. Dataset Description*

We used the HAM10000 dataset which contains 10015 images. From the dataset, we have disregarded 15 images to make data distribution perfect. The dimension of each image is 600×450. During the training phase, all the images were resized to 224×224. The dataset is downloaded from the Harvard Dataverse Website [12]. The number of benign images is 6705 and the number of malignant images is 3295. We have taken 5000 images from the benign category and applied data augmentation to the malignant category to make it 5000. For this augmentation, we used rotation, flip, and zoom-in augmentation parameters. We split the dataset into three portions: training set 60%, validation set 20%, and test set 20% of the total data which is shown in **Table 1**. The images were captured using a Dermatoscope at a magnification of 20X. **Figure 3**, shows some of the example images from the dataset. Data were annotated by experts.

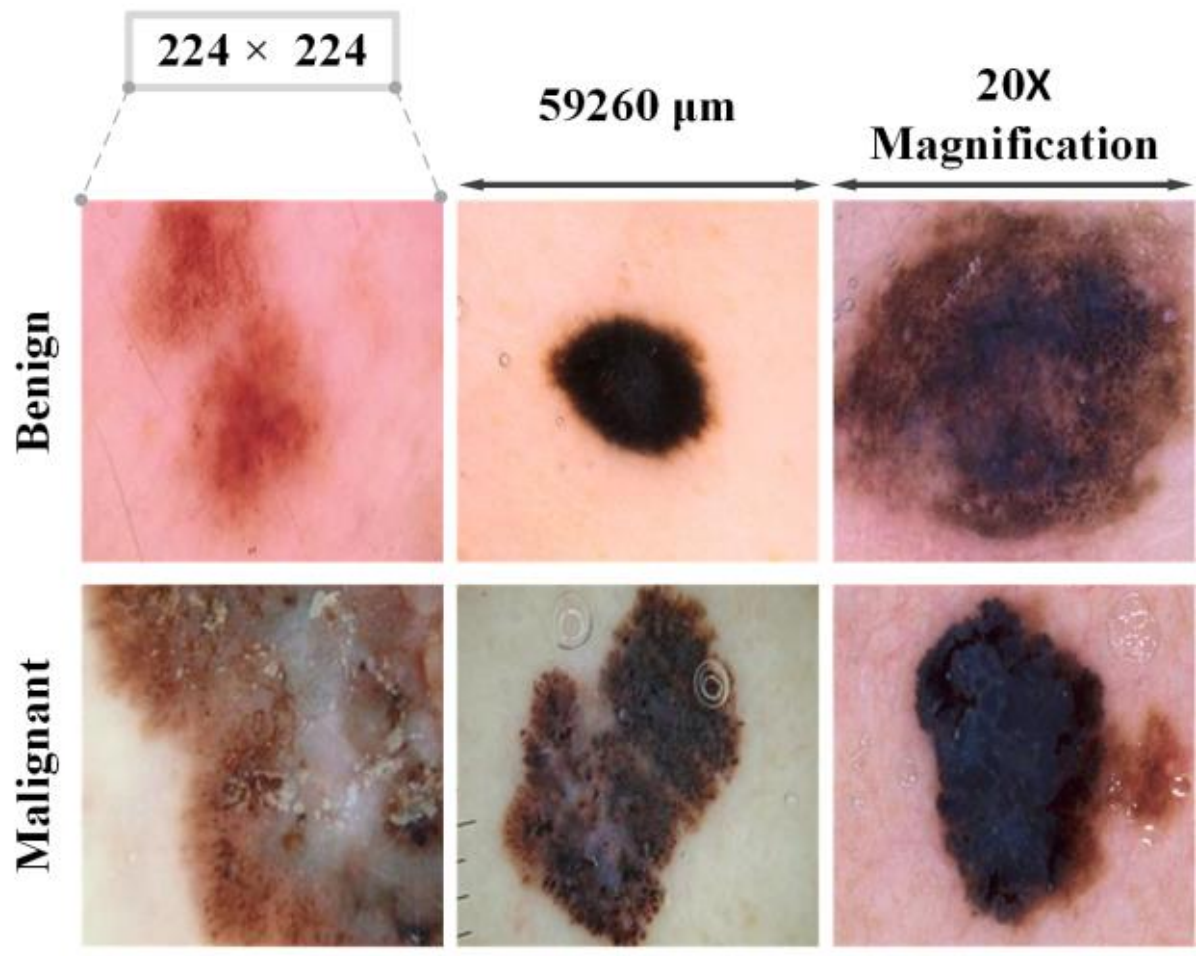

**Figure 3.** Example Images of Dataset

**Table 1.** Dataset Description

| Total images | Train images | Val images | Test images |
|---|---|---|---|
| Benign | 3000 | 1000 | 1000 |
| Malignant | 3000 | 1000 | 1000 |
| **Total** | **6000** | **2000** | **2000** |

### *3.2. Experimental Procedures*

The development of an artificial intelligence (AI) based digital skin cancer detection system that incorporates both segmentation and classification techniques is presented in this study. **Figure 4** illustrates the overall segmentation and classification process. The classification part of **Figure 4** is illustrated in detail in **Figure 5**. The system aims to improve the accuracy and efficiency of skin cancer diagnosis. We are proposing a method that will use the HAM10000 dataset to train the segmentation and classification models that can be used in digital smart devices to instantly classify skin cancers from new images. For the training task, all of the cancer images were segmented using binary masking where the white area represents the cancerous area and the black portion represents irrelevant areas (skin, background, etc.). After the training phase, the segmentation is generated that is capable of segmenting the cancer region automatically. Recently, Facebook's SAM model has gained huge popularity in terms of segmenting any objects almost perfectly but has not been employed in any biomedical imaging experiment officially. That is why we have trained our model using the SAM architecture and have achieved good results. After the segmentation has been done the segmented images need to be classified. Rather than employing new models we have experimented with some existing vision transformer architectures and compared their results to find out the best one for the classification task.

For the classification part, CNN-based network and Vision Transformer-based architecture can be used. Among the convolutional neural networks, there are many types such as RNN, ANN, etc. are very popular. However, the Vision Transformer architecture has some advantages such as ViT models can capture long-range dependencies in the input images more effectively than CNNs by utilizing the self-attention mechanism. ViT can learn contextual relationships between different patches, allowing for a better understanding of global image structures and improving performance on tasks that require modeling long-range dependencies, it can handle images of varying resolutions without the need for architectural modifications. CNNs, on the other hand, typically require resizing or cropping of images to match a specific input size. ViT's patch-based approach enables flexibility in processing images of different dimensions, making it easier to adapt to diverse datasets. ViT models tend to have fewer parameters compared to CNNs, which can lead to more computationally efficient training and inference. This efficiency is partly due to the absence of convolutional layers and shared weights across spatial locations. Consequently, ViT can be more memory-efficient and require fewer computational resources. The attention mechanism used in ViT provides interpretability advantages, as it allows the model to assign importance weights to different patches in the image. This attention mechanism enables a better understanding of which regions contribute more to the model's decision-making process, aiding in model interpretability and facilitating analysis of the learned representations. ViT models can benefit from transfer learning using pretraining on large-scale datasets. By pretraining on a large dataset, ViT can learn generic visual representations that can be fine-tuned on specific downstream tasks, even with limited task-specific labeled data. This transfer learning capability contributes to improved performance and efficiency. Considering these factors, we are proposing a Vision

Transformer for the classification task of skin cancer. We have tried the following six different ViT models: ViT-Google, ViT-MAE, ViT-ResNet50, ViT-VAN, ViT-BEiT, and ViT-DiT. Among these ViT-Google model outperformed the other ones.

A detailed workflow regarding acquiring cancer images, creating a segmentation model, creating a classification model, decision-making from new images, and data visualization for diagnosis is described in **Algorithm 1**.

**Algorithm 1:**

1. **Input**: $I_{DERMATOSCOPY}$, $SegNet_{SAM}$
   $I_{DERMATOSCOPY}$= the 20X Dermatoscopy images
   $SegNet_{SAM}$= the SAM segmentation
2. **Initialization:**
   i. $I$ = Resize $I_{DERMATOSCOPY}$ into 224 × 224-pixel
   ii. $I_{SEG}$ = Apply $SegNet_{SAM}$ on $I$
   iii. $I_{SEG}$ = Apply Binary Mask on $I_{SEG}$
3. **ViT:**
   i. **Input:** $I_{SEG}$
   ii. $I_{PATCH}$ = Convert $I_{SEG}$ into 14×14 patches
   iii. **While** $m$: = $I_{PATCH} \leq I_{PATCH}$ Count
      i. $I_{LP}[m] = I_{PATCHUNROLLED}[m] \times E\,[\,][\,]$
      ii. $D$ = ***Concatenate*** $(E_{POS}[m], I_{LP}[m])$
   iv. **End While**
   v. **While** $n <= L$ **do**
      i. $D'_n$ = **MSA** (**LN**($D_{n-1}$)) + $D_{n-1}$
      ii. $D_n$ = **MLP** (**LN** ($D'_n$)) + $D'_n$
   vi. **End While**
   vii. $I_{FINAL}$ = **LN**($D_L$)
4. **If** $I_{FINAL} \leq$ *Threshold Value* Then
   Result: = "**Benign**"
   **Else**
   Result: = "**Malignant**"
   **End If**
5. **Return** Result

In the primary phase, cancer images of 20x magnification are acquired by dermatoscopes. After resizing the images to 224x224 pixels they are given as input to the SAM automatically. After the SAM output is generated, it is converted to binary masking. The processed output images are given to the ViT model for classification. In the ViT model, it takes the input image which then is divided into fixed-size patches, which are then linearly transformed into lower-dimensional embeddings. Each patch serves as an input token to the Transformer. Positional embeddings are used to provide spatial information about the patches. Inside the positional embedding, *Sine* and *Cosine* functions of different frequencies are used to represent the relative positions of the patches. The core of the Vision Transformer is the Transformer encoder which consists of multiple layers of self-attention and feed-forward neural networks. The self-attention mechanism allows the model to capture global and local dependencies within the image, enabling it to attend to relevant patches and learn informative representations. This self-attention mechanism helps the network to identify necessary features. Every pre-trained model has its mapping criteria. In our experimental model, we have used Google's patch 32 version to extract the image feature and train the Vision Transformer. Within each Transformer layer, multi-head attention is applied to capture different types of relationships between patches. It allows the model to attend to different parts of the image simultaneously, enabling it to learn diverse and meaningful representations. In addition to the self-attention mechanism, each Transformer layer also contains feed-forward neural networks, which process the embeddings and facilitate the learning of non-linear relationships. Layer normalization is applied after each self-attention and feed-forward network, which helps stabilize the training process and improves the model's ability to generalize to different inputs. A classification head (described in **Figure 5**) is added on top of the Transformer encoder to produce class probabilities. We have used a fully connected layer followed by a SoftMax activation function. The Vision Transformer we employed uses supervised learning on large-scale labeled image datasets. In the final phase, the image is classified whether it is benign or malignant. Our experimental model successfully differentiates cancerous cells from non-cancerous ones. Then the result will be visualized in the electronic device for analysis.

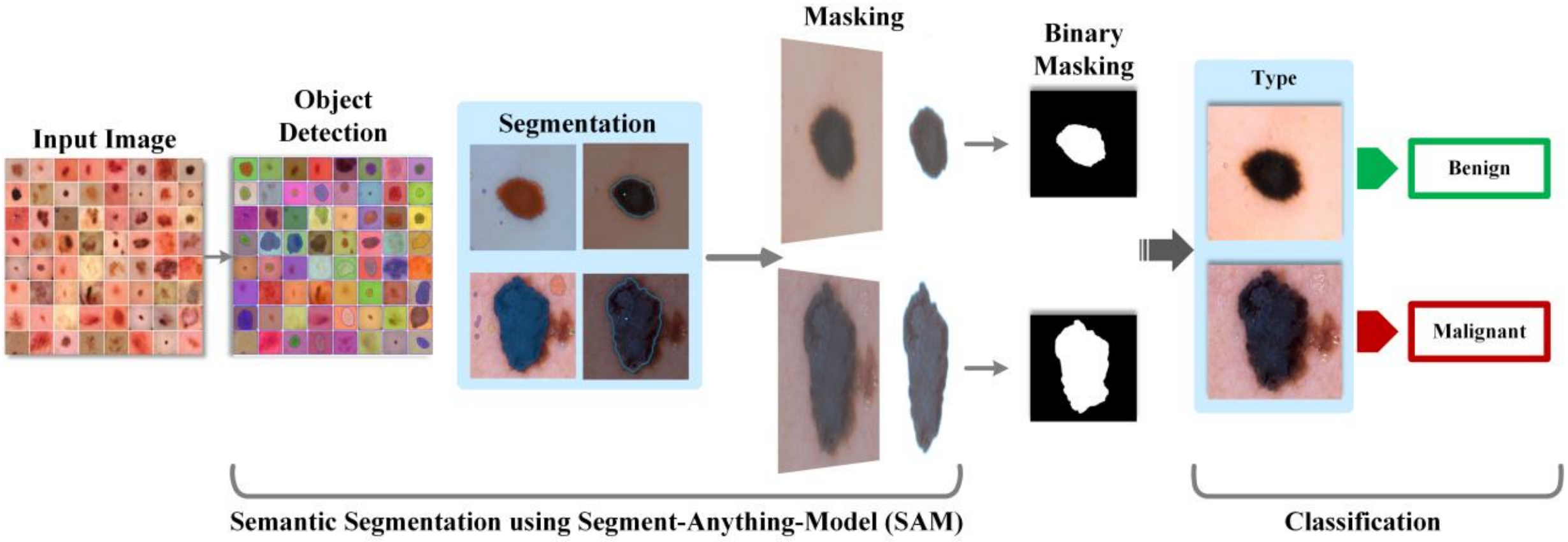

**Figure 4.** Step-by-Step Processes for Cancer Segmentation and Classification

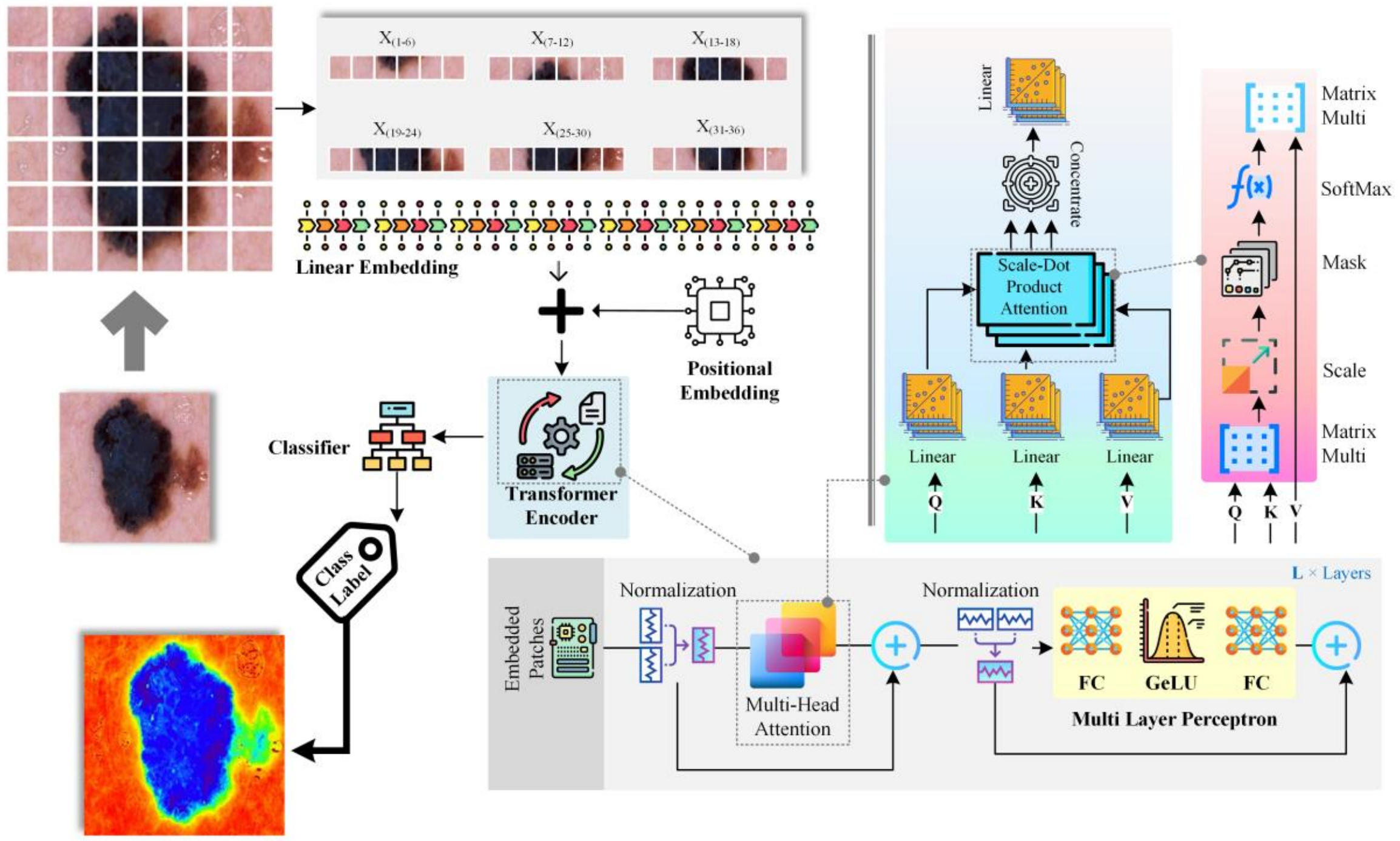

**Figure 5.** Vision Transformer-based Skin Cancer Classification Model.

### 3.3. Experimental Setup

For our experiment, we used both 'TensorFlow' [19] and 'PyTorch' [20]. We have utilized the TensorFlow-based interface to employ the SAM for the segmentation process. For running the segmentation experiment we have used the following hyperparameters mentioned in **Table 2**.

**Table 2.** Hyperparameter Optimization for Segmentation

| Hyperparameter | Optimization Space |
|---|---|
| Epoch | 100 |
| Batch size | 32 |
| Learning rate | 0.0001 |
| Optimizer | 'RMSprop' |
| Loss function | [Dice-Coefficient Loss] |

We have utilized the PyTorch-based interface to employ the Vision Transformers. This environment was set up in a different computer to avoid dependency mismatch. The hyperparameter optimization was done as mentioned in **Table 3**.

**Table 3.** Hyperparameter Optimization for Classification using ViT

| Hyperparameter | Optimization Space |
|---|---|
| Epoch | 100 |
| Batch size | 32 |
| Learning rate | 0.00002 |
| Optimizer | 'Adam' |
| Loss function | [Cross Entropy Loss] |

Our experiment was done by using two different computers; one of them running on Windows 11 OS and equipped with an NVIDIA GeForce RTX 3090(24 GB) GPU and 32GB RAM another was running on Windows 11 OS and equipped with an NVIDIA GeForce RTX 3080Ti (16 GB) GPU and 64GB RAM.

## 4. Results & Discussion

### *4.1. Accuracy*

**Figure 7, Figure 8,** and **Figure 9** illustrate the accuracy, Intersection Oven Union (IOU), and Dice-Coefficient curves respectively for the Segment Anything Model (SAM). **Table 4** shows the training accuracy, IOU, and Dice-Coefficient as well as the validation accuracy, IOU, and Dice-Coefficient for the segmentation model**.** **Figure 10** depicts the cancer segmentation result by the Segment Anything Model (SAM). **Table 5** Provides the accuracy & and loss scores of different ViT classifiers. **Figure 11** illustrates the accuracy and loss curve for various models tested for ViT.

#### *4.1.1. Segmentation*

Pixel accuracy is a commonly used metric for evaluating the performance of image segmentation algorithms. It measures the exactness of the segmentation results by comparing the predicted segmentation mask to the ground truth mask at the pixel level. Pixel accuracy calculates the percentage of correctly classified pixels in the segmentation output, indicating how well the algorithm accurately assigns each pixel to its correct class or region. It is calculated by dividing the number of correctly classified pixels by the total number of pixels in the image. Pixel accuracy is a straightforward metric that provides a basic measure of segmentation performance. However, it does not consider the distinction between different classes or the spatial relationships between pixels. As a result, it may not provide a comprehensive assessment of segmentation quality in cases where certain classes are more critical than others or when precise boundaries are important. While pixel accuracy is useful for obtaining a general understanding of the segmentation accuracy, it is often accompanied by additional evaluation metrics, such as Intersection over Union (IoU) or Dice coefficient, to provide a more comprehensive assessment. These metrics take into account the overlap between the predicted and ground truth masks, providing a more nuanced evaluation of segmentation performance.
IoU calculates the ratio of the intersection area between the predicted segmentation mask and the ground truth mask to the union area of both masks. It provides a measure of overlap or similarity between the predicted and ground truth masks, taking into account both true positive and false positive predictions. IoU is more informative than pixel accuracy as it considers the spatial relationship and overlap between the segmented regions. **Equation (1)** represents the IoU formula. **Figure 6** demonstrates the graphical representation of IOU.

$$IoU = \frac{Intersection}{Union} \qquad \ldots\ldots\ldots\ldots\ldots (1)$$

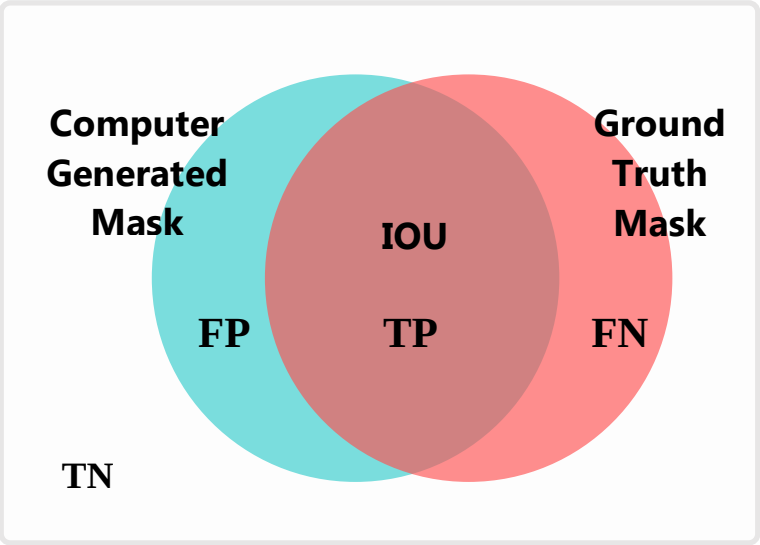

**Figure 6.** Intersection Over Union (IOU). Here, FP=False Positive, TP=True Positive, FN=False Negative, TN=True Negative.

The Dice coefficient measures the similarity between the predicted and ground truth masks by computing the ratio of twice the intersection area to the sum of areas of both masks. It balances precision and recall, providing a measure of overall segmentation performance. Like IoU, the Dice coefficient captures the spatial overlap between segmented regions and is more informative than pixel accuracy. **Equation (2)** is representing the Dice-Coefficient formula.

$$Dice - Coefficient = \frac{2 \times Intersection}{Union + Intersection} \qquad \ldots \ldots \ldots \ldots \ldots (2)$$

IoU is more intuitive and easier to interpret. It represents the ratio of the intersection area between the predicted and ground truth masks to the union area of both masks. It directly reflects the overlap or similarity between the two masks, making it easier to understand the quality of segmentation in terms of spatial overlap. That is why we prefer the IoU matrix over other matrices in terms of measuring segmentation performance.

In our experiment, while applying the SAM model we have gotten over 96% of areas matched on average under the IoU matrix. In **Figure 10**, we can see a side-by-side comparison of two example images which indicates the IoU accuracy.

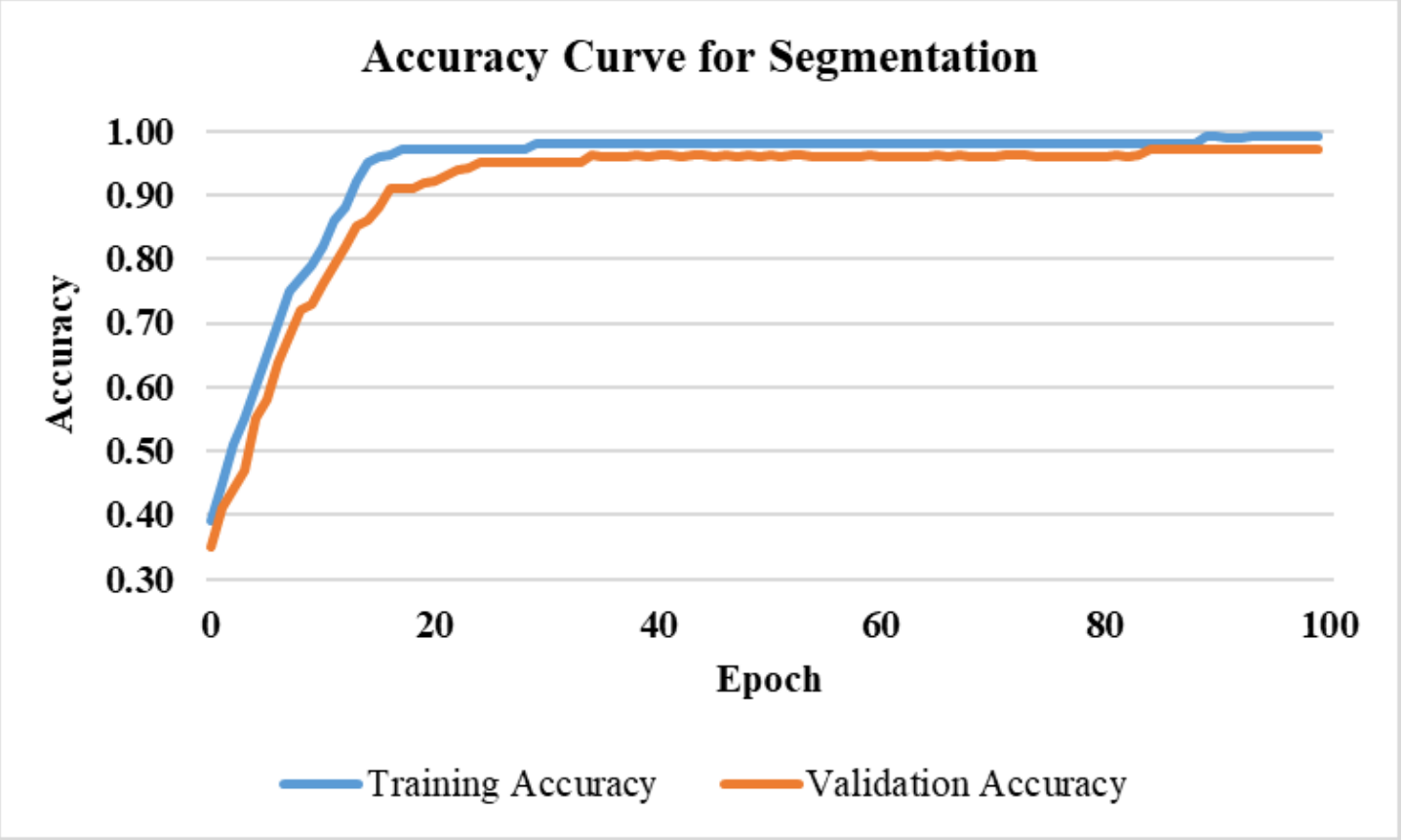

**Figure 7.** Accuracy Curve for Segmentation

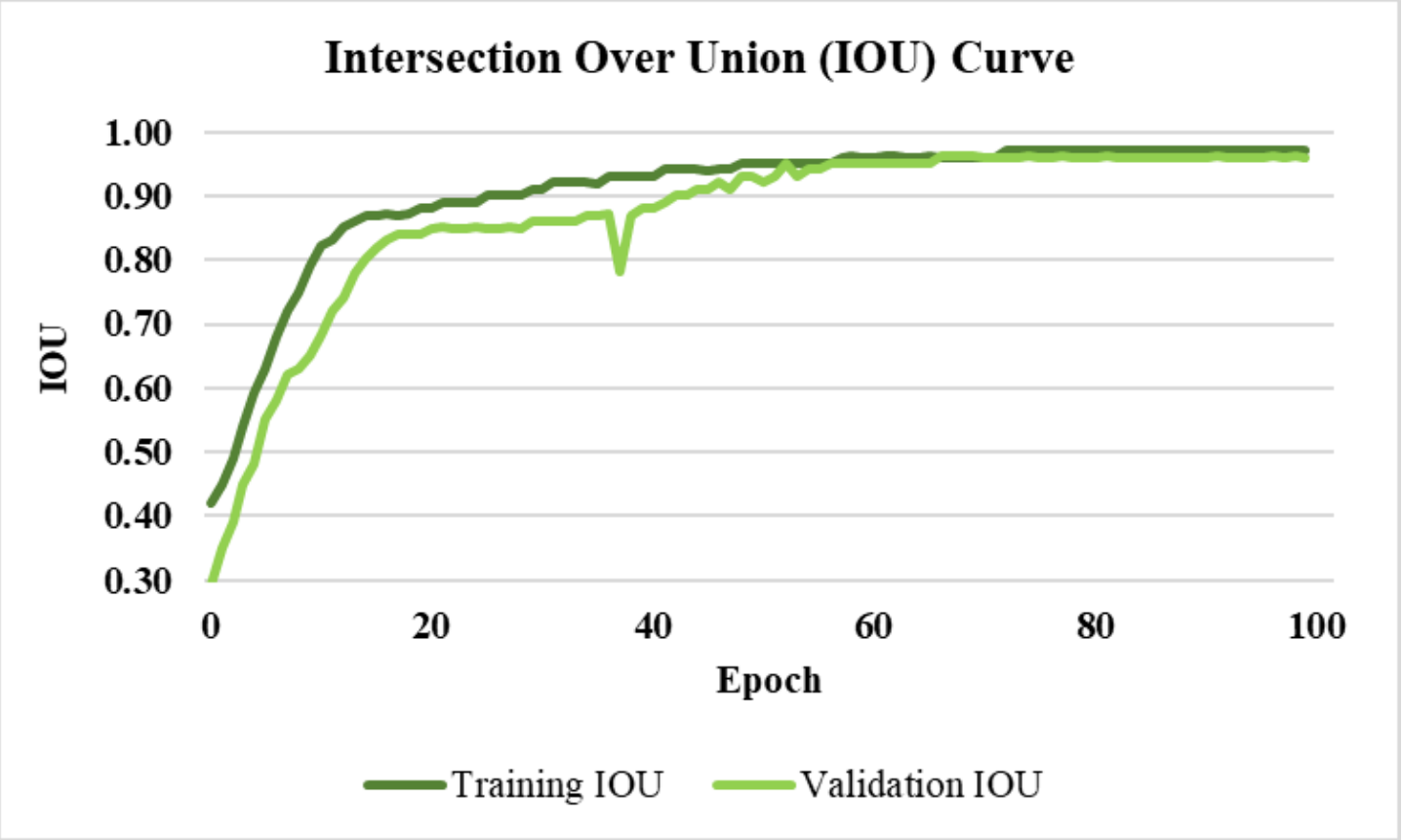

**Figure 8.** Intersection Oven Union (IOU) Curve for Segmentation

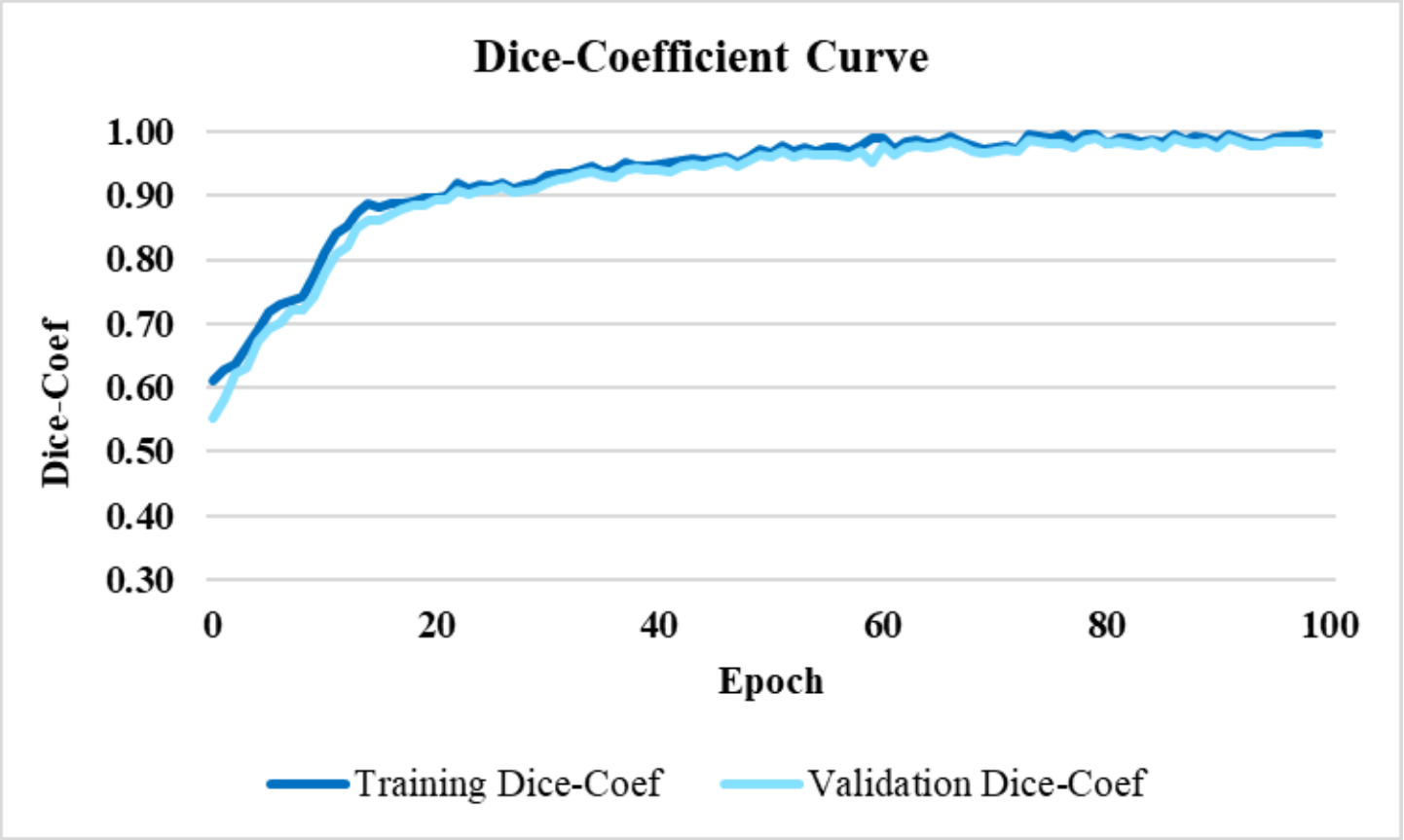

**Figure 9.** Dice-Coefficient Curve for Segmentation

**Table 4**. Training and validation accuracy, IOU & Dice-Coefficient

| | Accuracy | IOU | Dice-Coefficient |
|---|---|---|---|
| **Training** | 0.99057 | 0.97043 | 0.99422 |
| **Validation** | 0.97129 | **0.96071** | 0.98139 |

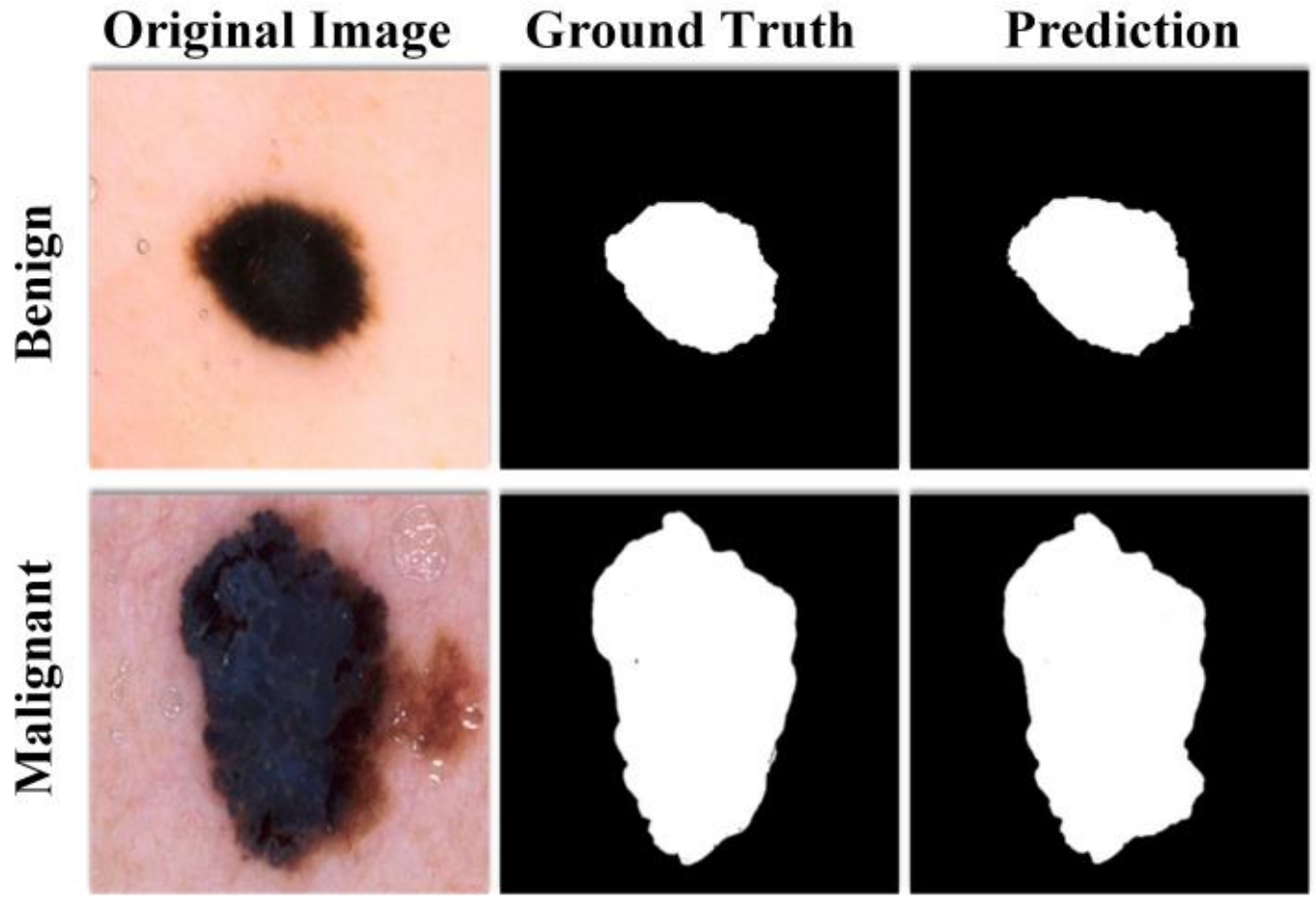

**Figure 10.** Cancer Segmentation Result by Segment Anything Model (SAM)

*4.1.2. Classification*

Using the same dataset, we have applied 6 different pre-trained models to employ our Vision Transformer-based classification. The obtained accuracy for the following ViT-Google, ViT-MAE, ViT-ResNet50, ViT-VAN, ViT-BEiT, and ViT-DiT models are mentioned in detail in **Table 5**. Comparing the accuracies, we have used Google's ViT (patch 32) model to be employed in the overall skin cancer detection smart system. **Figure 11**. Illustrates the accuracy difference between ViT-Google and the other models. Another thing can be also observed from **Figure 11.** that the loss scores are significantly better than the other five models.

**Table 5.** Accuracy and loss scores of different ViT classifiers

| Model | Training Accuracy | Training Loss | Validation Accuracy | Validation Loss | Test Accuracy |
|---|---|---|---|---|---|
| ViT-Google | 0.99375 | 0.40280 | 0.96275 | 0.50341 | **0.96150** |
| ViT-MAE | 0.96975 | 0.77760 | 0.92500 | 0.96414 | 0.92150 |
| ViT-ResNet50 | 0.87175 | 2.45290 | 0.81000 | 2.45290 | 0.80900 |
| ViT-VAN | 0.95825 | 0.95315 | 0.92500 | 1.04780 | 0.92400 |
| ViT-BEiT | 0.92500 | 1.49015 | 0.87500 | 1.89523 | 0.86250 |
| ViT-DiT | 0.86175 | 1.54550 | 0.82500 | 0.82500 | 0.81900 |

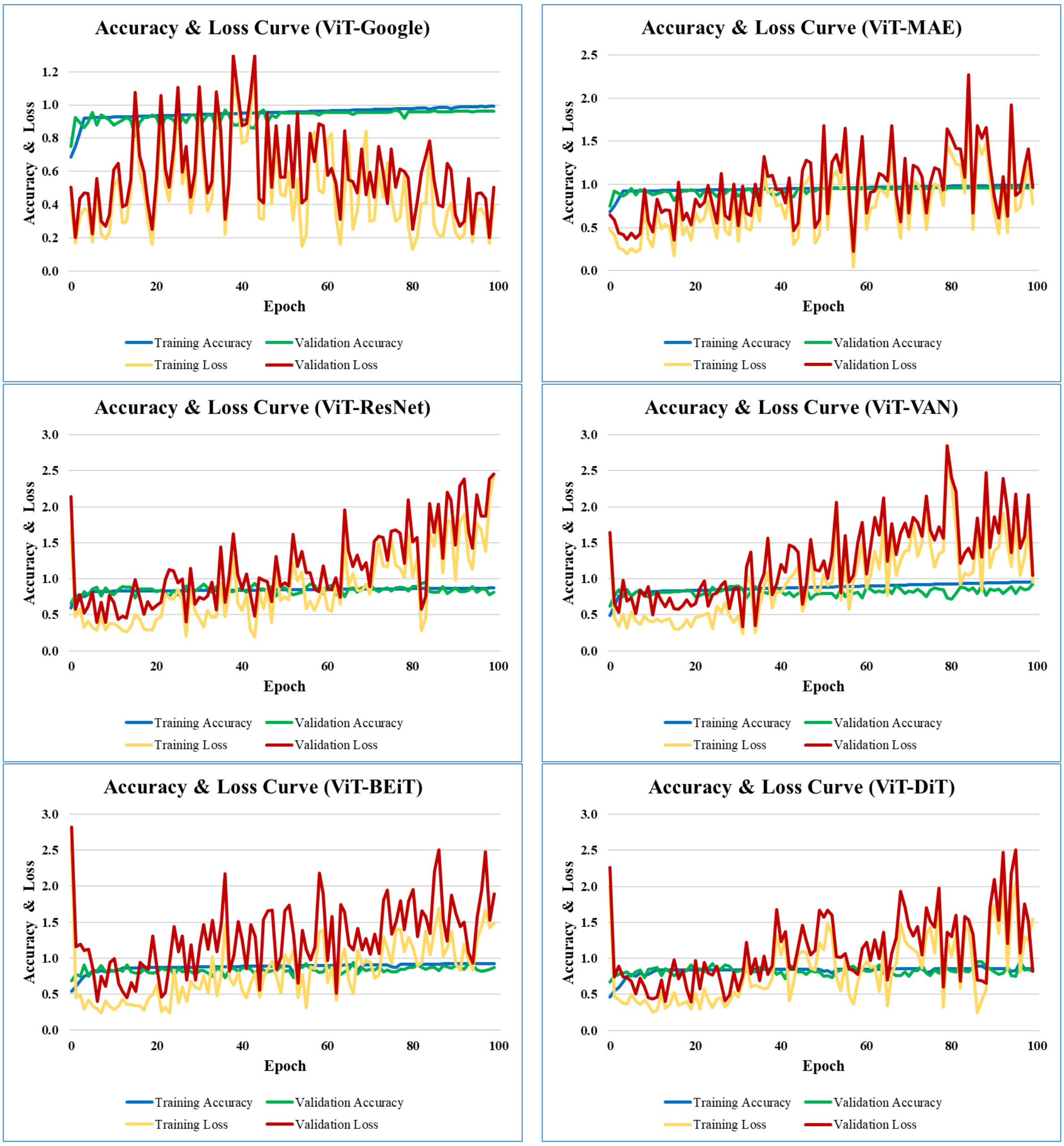

**Figure 11.** Accuracy Curve for various Models tested for ViT

The F-1 score is a widely used metric in classification evaluations that assesses the overall effectiveness of a model in binary or multi-class classification tasks. It provides a balanced measure of accuracy by calculating the harmonic mean of precision and recall. The precision determines the proportion of correct positive predictions out of all positive predictions made by the model, reflecting its ability to accurately identify positive instances. On the other hand, recall, also known as sensitivity or true positive rate, measures the proportion of true positive predictions out of all actual positive instances in the dataset, indicating the model's capability to capture all positive instances. By combining precision and recall into a single metric, the F-1 score offers a balanced assessment of a model's performance. It is particularly valuable when dealing with imbalanced datasets or when both precision and recall hold equal importance. The F-1 score ranges from 0 to 1, where a value of 1 signifies perfect precision and recall, while 0 represents the poorest performance. The formulas for precision, recall, F-1 score, and accuracy are provided in equations (3), (4), (5), and (6) respectively. The classification report of the transfer learning models can be found in **Table 6**.

$$\text{Precision} = \frac{\text{True Positive}}{\text{True Positive+False Positive}} \quad \ldots\ldots\ldots(3)$$

$$\text{Recall} = \frac{\text{True Positive}}{\text{True Positive+False Negative}} \quad \ldots\ldots\ldots(4)$$

$$\text{F1 Score} = 2 * \frac{\text{Precision}\times\text{Recall}}{\text{Precision}+\text{Recall}} \quad \ldots\ldots\ldots(5)$$

$$\text{Accuracy} = \frac{\text{True Negative}+\text{True Positive}}{\text{True Negative}+\text{False Positive}+\text{True Positive}+\text{False Negative}} \quad \ldots\ldots\ldots(6)$$

**Table 6.** Classification Report of the best combination for each transfer learning model

| Model | | Precision | Recall | f1-score | Model | | Precision | Recall | f1-score |
|---|---|---|---|---|---|---|---|---|---|
| ViT (Google) | Benign | 0.96950 | 0.95300 | 0.96120 | ViT (VAN) | Benign | 0.93620 | 0.91000 | 0.92290 |
| | Malignant | 0.95380 | 0.97000 | 0.96180 | | Malignant | 0.91250 | 0.93800 | 0.92500 |
| | **Accuracy** | | 0.96150 | | | **Accuracy** | | 0.92400 | |
| | **Macro-F1** | | 0.96150 | | | **Macro-F1** | | 0.92400 | |
| | **Weighted-F1** | | 0.96150 | | | **Weighted-F1** | | 0.92400 | |
| ViT (Mae) | Benign | 0.94230 | 0.89800 | 0.91960 | ViT (BEiT) | Benign | 0.91430 | 0.80000 | 0.85330 |
| | Malignant | 0.90260 | 0.94500 | 0.92330 | | Malignant | 0.82220 | 0.92500 | 0.87060 |
| | **Accuracy** | | 0.92150 | | | **Accuracy** | | 0.86250 | |
| | **Macro-F1** | | 0.92150 | | | **Macro-F1** | | 0.86200 | |
| | **Weighted-F1** | | 0.92150 | | | **Weighted-F1** | | 0.86200 | |
| ViT (ResNet50) | Benign | 0.81470 | 0.80000 | 0.80730 | ViT (DiT) | Benign | 0.83160 | 0.80000 | 0.81550 |
| | Malignant | 0.80350 | 0.81800 | 0.81070 | | Malignant | 0.80730 | 0.83800 | 0.82240 |
| | **Accuracy** | | 0.80900 | | | **Accuracy** | | 0.81900 | |
| | **Macro-F1** | | 0.80900 | | | **Macro-F1** | | 0.81890 | |
| | **Weighted-F1** | | 0.80900 | | | **Weighted-F1** | | 0.81890 | |

In **Figure 12**. the confusion matrices of the tested models are presented. From the confusion matrices, we can observe that in the misclassification cases, the false negative rate is greater than false positive cases for all models while the overall test accuracy of ViT Google greater than those of others.

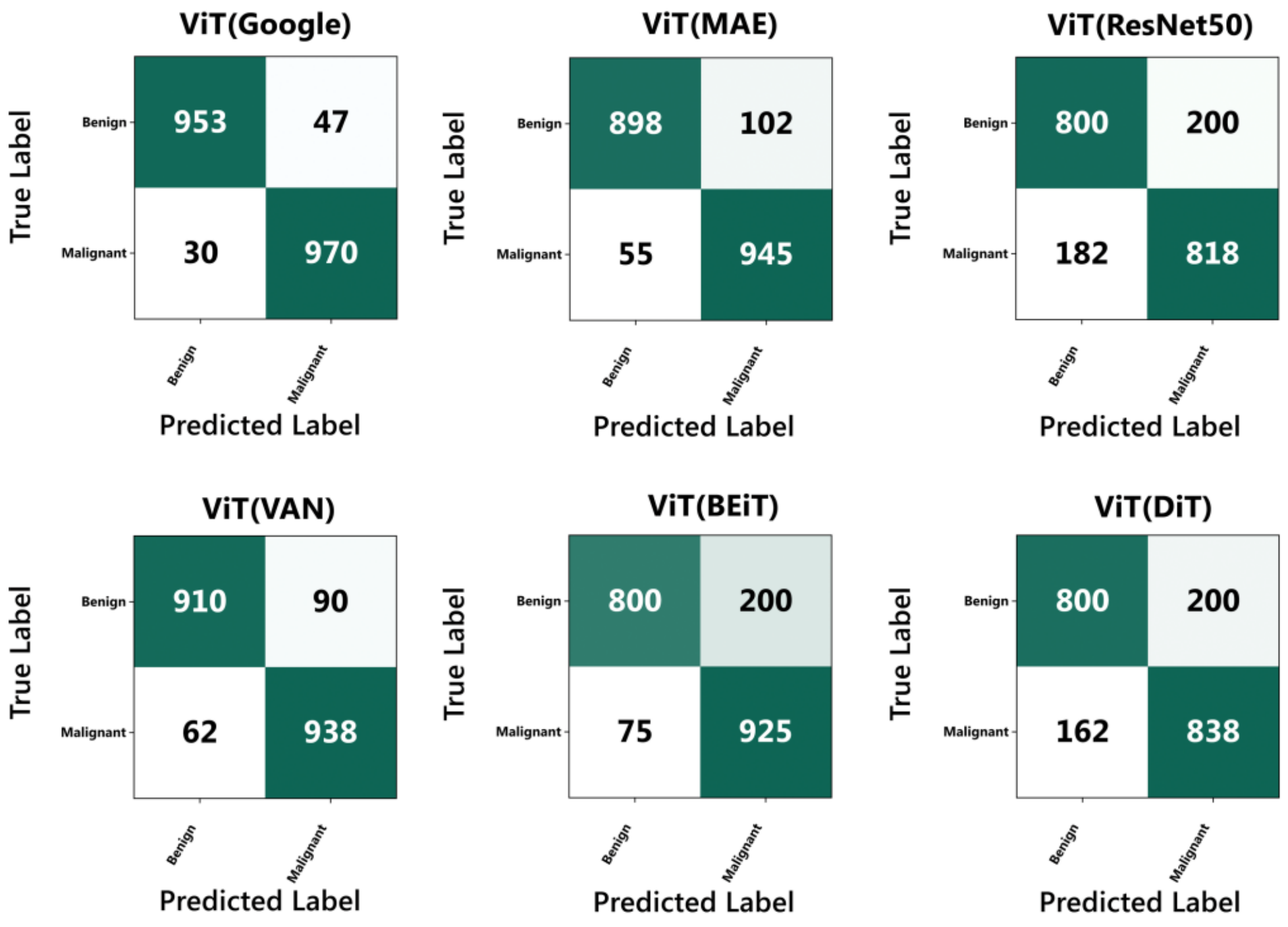

**Figure 12.** Confusion Matrices

*4.2. Accuracy Comparison*

**Figure 13**. compares the training, validation, and test accuracy of the applied models. **Table 7.** Describes the value difference between training accuracy and test accuracy of all tested models. The value difference is an indication of training performance. The lower the value is the better the training process is. However, one thing should be taken into consideration this value may be low even if the test accuracy is low. In that case, this value cannot be a measurement factor regarding assessing the training performance. From Table 7. It is clear that for the ViT-Google model, the value difference between training accuracy and test accuracy is low at the same time the test accuracy is higher than the other models. Since the ViT-Google model has performed better we have chosen that model for final implementation. For this reason, we have applied 5-fold cross-validation on that model. **Figure 14**. Describe the 5-fold cross-validation for the ViT-Google model. The average validation accuracy found using the 5-fold cross-validation method is around 96.15%. The Receiver Operating Characteristic (ROC) curve is a widely used evaluation tool in medical research. It provides a graphical representation of the trade-off between the true positive rate (sensitivity) and false positive rate (1-specificity) for different classification thresholds. The area under the ROC curve (AUC) summarizes the overall performance of a diagnostic or predictive model. In this study, the ROC analysis was employed to assess the discrimination ability of our selected model for skin cancer classification. The AUC value serves as a quantitative measure of the model's ability to distinguish between cancerous and non-cancerous cases, with higher values indicating better performance. From **Figure 15**. We can see that the ROC score or AUC of ViT-Google is 99.49%.

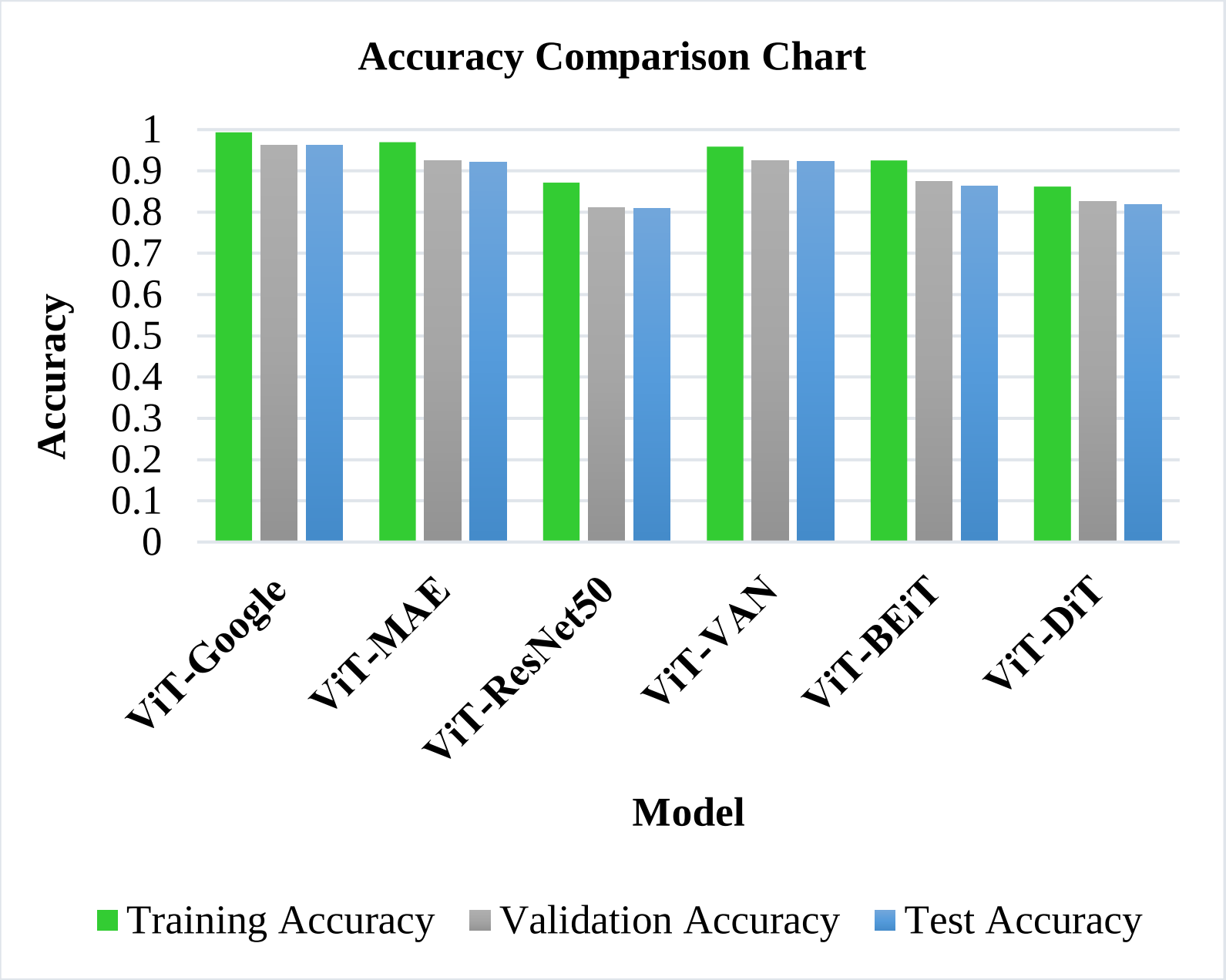

**Figure 13.** Accuracy Comparison Chart

**Table 7.** Accuracy Difference

| Model | Training Accuracy | Test Accuracy | Accuracy Difference |
|---|---|---|---|
| ViT-Google | 0.99375 | **0.9615** | ***0.03225*** |
| ViT-MAE | 0.96975 | 0.9215 | *0.04825* |
| ViT-ResNet50 | 0.87175 | 0.8090 | *0.06275* |
| ViT-VAN | 0.95825 | 0.9240 | *0.03425* |
| ViT-BEiT | 0.92500 | 0.8625 | *0.06250* |
| ViT-DiT | 0.86175 | 0.8190 | *0.04275* |

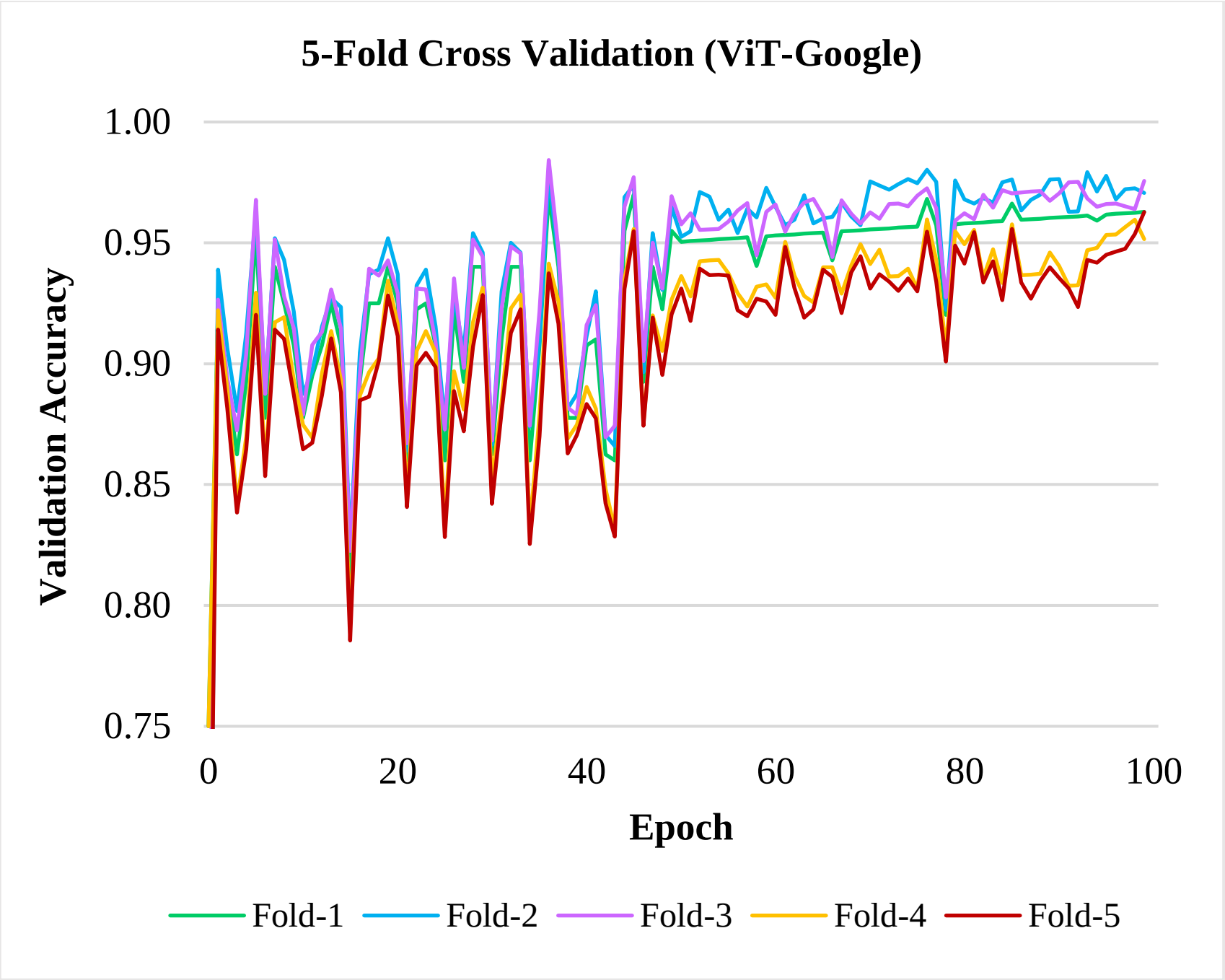

**Figure 14.** 5-Fold Cross Validation for ViT-Google

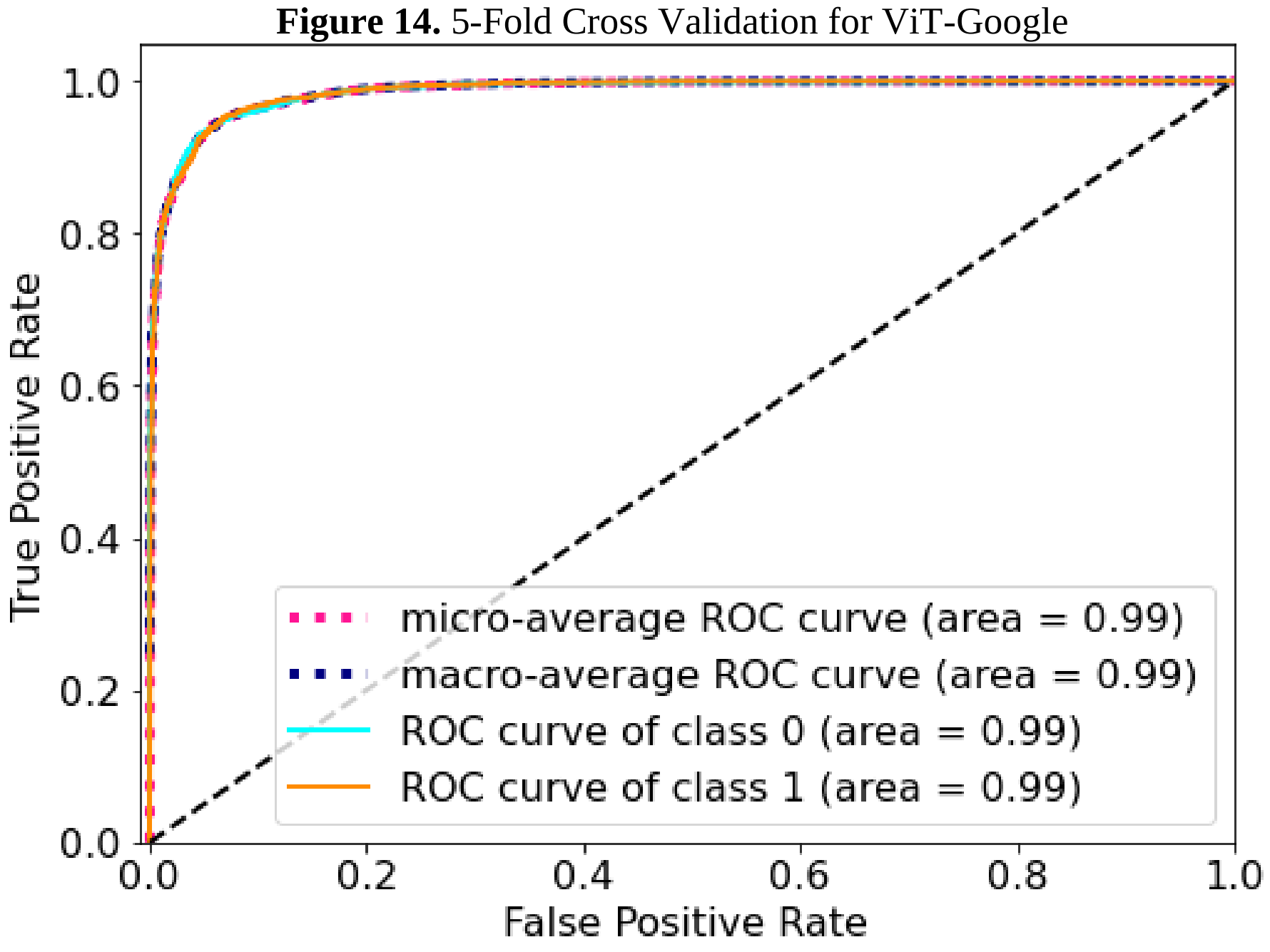

**Figure 15.** Receiver Operating Characteristics Curve for ViT-Google

## 5. Conclusion

In this research, we utilize Google's 32-patch-based Vision Transformer (ViT) model to address the identification of skin cancer. The prevalence of skin cancer has increased significantly worldwide due to the depletion of the Ozone layer, posing a substantial threat to communities. Delayed testing procedures, limited facilities, and a lack of early-stage diagnosis have led to numerous deaths. To solve this, intelligent smart technologies are necessary to establish rapid and effective testing procedures. Our experimental approach involves employing the ViT model in conjunction with the SAM segmentation model for skin cancer identification and classification. The application of the Segment Anything Model (SAM) enables the segmentation of cancerous regions in the images, yielding an Intersection over Union (IOU) of 96.01% and a Dice Coefficient of 98.14%. Also, our classification experiments with Google's ViT model yield impressive results, achieving 96.15% accuracy and 99.49% ROC score.

Consequently, we conclude that the ViT-Google (patch 32) model can be trained as an efficient skin cancer detection model using the Vision Transformer and implemented in smart devices for swift detection, thereby providing valuable support to pathologists. However, it is essential to acknowledge that our ViT model-based method is more suitable for fair-skinned individuals due to the dataset acquired from Harvard University in America, where the predominant population is white. Our future work will expand the dataset to encompass a more diverse range of cases from individuals of various ethnic backgrounds. Additionally, federated learning can be employed to update the existing model with new data, allowing for iterative improvements without retraining the entire model.

## Reference

1. National Cancer Institute (2018). *Common Cancer Sites - Cancer Stat Facts*. [online] SEER. Available at: https://seer.cancer.gov/statfacts/html/common.html.
2. National Cancer Institute (2018). *Melanoma of the Skin - Cancer Stat Facts*. [online] SEER. Available at: https://seer.cancer.gov/statfacts/html/melan.html.
3. *Cancer Facts and Figures 2023*. American Cancer Society. https://www.cancer.org/content/dam/cancer-org/research/cancer-facts-and-statistics/annual-cancer-facts-and-figures/2023/2023-cancer-facts-and-figures.pdf. *Accessed January 12, 2023.*
4. Global Burden of Disease Cancer Collaboration. Global, regional, and national cancer incidence, mortality, years of life lost, years lived with disability, and disability-adjusted life-years for 29 cancer groups, 1990 to 2017. *JAMA Oncol*. 2019;5(12):1749-1768. doi:10.1001/jamaoncol.2019.2996.
5. Wehner MR, Chren MM, Nameth D, et al. International prevalence of indoor tanning: a systematic review and meta-analysis. *JAMA Dermatol* 2014; 150(4):390-400. doi:10.1001/jamadermatol.2013.6896.
6. Stern, RS. Prevalence of a history of skin cancer in 2007: results of an incidence-based model. *Arch Dermatol* 2010; 146(3):279-282.
7. Gloster HM, Neal K. Skin cancer in skin of color. *J Am Acad Dermatol* 2006; 55:741-60.
8. Bradford, Porcia T. Skin Cancer in Skin of Color. *Dermatol Nurs* 2009 Jul-Aug; 21(4): 170-178.
9. Han D, Zager JS, Han G, et al. The unique clinical characteristics of melanoma diagnosed in children. Ann Surg Oncol. 2012;19(12):3888–3895. doi:10.1245/s10434-012-2554-5
10. Cancer Research UK (2019). *Risks and causes | Skin cancer | Cancer Research UK*. [online] Cancerresearchuk.org. Available at: https://www.cancerresearchuk.org/about-cancer/skin-cancer/risks-causes.
11. dermnetnz.org. (n.d.). *ABCDEFG of melanoma | DermNet NZ*. [online] Available at: https://dermnetnz.org/topics/abcdes-of-melanoma.
12. Tschandl, P., Rosendahl, C. & Kittler, H. The HAM10000 dataset is a large collection of multi-source dermatoscopic images of common pigmented skin lesions. *Sci Data* **5**, 180161 (2018). https://doi.org/10.1038/sdata.2018.161
13. Kirillov A, Mintun E, Ravi N, et al (2023) Segment Anything. arXiv:230402643 [cs]
14. Dosovitskiy, Alexey, et al. "An image is worth 16x16 words: Transformers for image recognition at scale." *arXiv preprint arXiv:2010.11929* (2020).
15. Guo, Meng-Hao, et al. "Visual attention network." *arXiv preprint arXiv:2202.09741* (2022).
16. He, K., Chen, X., Xie, S., Li, Y., Dollár, P. and Girshick, R. (2021). Masked Autoencoders Are Scalable Vision Learners. *arXiv:2111.06377 [cs]*. [online] Available at: https://arxiv.org/abs/2111.06377v2.
17. Li, J., Xu, Y., Tengchao Lv, Cui, L., Zhang, C. and Wei, F. (2022). DiT: Self-supervised Pre-training for Document Image Transformer. doi:https://doi.org/10.1145/3503161.3547911.
18. Bao, H., Dong, L. and Wei, F. (2021). BEiT: BERT Pre-Training of Image Transformers. *arXiv:2106.08254 [cs]*. [online] Available at: https://arxiv.org/abs/2106.08254.
19. Google (2019). *TensorFlow*. [online] TensorFlow. Available at: https://www.tensorflow.org/.
20. PyTorch (2023). *PyTorch*. [online] Pytorch.org. Available at: https://pytorch.org/.
21. Dorj, UO., Lee, KK., Choi, JY. et al. The skin cancer classification using deep convolutional neural network. Multimed Tools Appl 77, 9909–9924 (2018). https://doi.org/10.1007/s11042-018-5714-1
22. Rezvantalab, A., Safigholi, H. and Karimijeshni, S. (2018). Dermatologist Level Dermoscopy Skin Cancer Classification Using Different Deep Learning Convolutional Neural Networks Algorithms. arXiv:1810.10348 [cs, stat]. [online] Available at: https://arxiv.org/abs/1810.10348."
23. K. M. Hosny, M. A. Kassem and M. M. Foaud, "Skin Cancer Classification using Deep Learning and Transfer Learning," 2018 9th Cairo International Biomedical Engineering Conference (CIBEC), Cairo, Egypt, 2018, pp. 90-93, doi: 10.1109/CIBEC.2018.8641762.
24. Dascalu, A. and David, E.O. (2019). Skin cancer detection by deep learning and sound analysis algorithms: A prospective clinical study of an elementary dermoscope. EBioMedicine, 43, pp.107–113. doi:https://doi.org/10.1016/j.ebiom.2019.04.055."
25. T. C. Pham, G. S. Tran, T. P. Nghiem, A. Doucet, C. M. Luong and V. -D. Hoang, "A Comparative Study for Classification of Skin Cancer," 2019 International Conference on System Science and Engineering (ICSSE), Dong Hoi, Vietnam, 2019, pp. 267-272, doi: 10.1109/ICSSE.2019.8823124.
26. Hekler, A., Utikal, J.S., Enk, A.H., Hauschild, A., Weichenthal, M., Maron, R.C., Berking, C., Haferkamp, S., Klode, J., Schadendorf, D., Schilling, B., Holland-Letz, T., Izar, B., von Kalle, C., Fröhling, S., Brinker, T.J., Schmitt, L., Peitsch, W.K., Hoffmann, F. and Becker, J.C. (2019). Superior skin cancer classification by the combination of human and artificial intelligence. European Journal of Cancer, [online] 120, pp.114–121. doi:https://doi.org/10.1016/j.ejca.2019.07.019.

27. T. Emara, H. M. Afify, F. H. Ismail and A. E. Hassanien, "A Modified Inception-v4 for Imbalanced Skin Cancer Classification Dataset," 2019 14th International Conference on Computer Engineering and Systems (ICCES), Cairo, Egypt, 2019, pp. 28-33, doi: 10.1109/ICCES48960.2019.9068110.
28. Chaturvedi, S.S., Gupta, K., Prasad, P.S. (2021). Skin Lesion Analyser: An Efficient Seven-Way Multi-class Skin Cancer Classification Using MobileNet. In: Hassanien, A., Bhatnagar, R., Darwish, A. (eds) Advanced Machine Learning Technologies and Applications. AMLTA 2020. Advances in Intelligent Systems and Computing, vol 1141. Springer, Singapore. https://doi.org/10.1007/978-981-15-3383-9_15
29. Mohapatra, S., Abhishek, N.V.S., Bardhan, D., Ghosh, A.A., Mohanty, S. (2021). Skin Cancer Classification Using Convolution Neural Networks. In: Tripathy, A., Sarkar, M., Sahoo, J., Li, KC., Chinara, S. (eds) Advances in Distributed Computing and Machine Learning. Lecture Notes in Networks and Systems, vol 127. Springer, Singapore. https://doi.org/10.1007/978-981-15-4218-3_42
30. Chen, M., Chen, W., Chen, W., Cai, L. and Chai, G. (2020). Skin Cancer Classification with Deep Convolutional Neural Networks. Journal of Medical Imaging and Health Informatics, 10(7), pp.1707–1713. doi:https://doi.org/10.1166/jmihi.2020.3078.
31. Jinnai, S.; Yamazaki, N.; Hirano, Y.; Sugawara, Y.; Ohe, Y.; Hamamoto, R. The Development of a Skin Cancer Classification System for Pigmented Skin Lesions Using Deep Learning. Biomolecules 2020, 10, 1123. https://doi.org/10.3390/biom10081123
32. Chaturvedi, S.S., Tembhurne, J.V. & Diwan, T. A multi-class skin Cancer classification using deep convolutional neural networks. Multimed Tools Appl 79, 28477–28498 (2020). https://doi.org/10.1007/s11042-020-09388-2
33. Garg, R., Maheshwari, S., Shukla, A. (2021). Decision Support System for Detection and Classification of Skin Cancer Using CNN. In: Sharma, M.K., Dhaka, V.S., Perumal, T., Dey, N., Tavares, J.M.R.S. (eds) Innovations in Computational Intelligence and Computer Vision. Advances in Intelligent Systems and Computing, vol 1189. Springer, Singapore. https://doi.org/10.1007/978-981-15-6067-5_65
34. Benedetti, P., Perri, D., Simonetti, M., Gervasi, O., Reali, G., Femminella, M. (2020). Skin Cancer Classification Using Inception Network and Transfer Learning. In: , et al. Computational Science and Its Applications – ICCSA 2020. ICCSA 2020. Lecture Notes in Computer Science(), vol 12249. Springer, Cham. https://doi.org/10.1007/978-3-030-58799-4_39
35. N. Gouda and J. Amudha, "Skin Cancer Classification using ResNet," 2020 IEEE 5th International Conference on Computing Communication and Automation (ICCCA), Greater Noida, India, 2020, pp. 536-541, doi: 10.1109/ICCCA49541.2020.9250855.
36. Ismail, M.A., Hameed, N., Clos, J. (2021). Deep Learning-Based Algorithm for Skin Cancer Classification. In: Kaiser, M.S., Bandyopadhyay, A., Mahmud, M., Ray, K. (eds) Proceedings of International Conference on Trends in Computational and Cognitive Engineering. Advances in Intelligent Systems and Computing, vol 1309. Springer, Singapore. https://doi.org/10.1007/978-981-33-4673-4_58
37. H. K. Kondaveeti and P. Edupuganti, "Skin Cancer Classification using Transfer Learning," 2020 IEEE International Conference on Advent Trends in Multidisciplinary Research and Innovation (ICATMRI), Buldhana, India, 2020, pp. 1-4, doi: 10.1109/ICATMRI51801.2020.9398388.
38. Maiti, Ananjan, et al. "Skin Cancer Classification Through Quantized Color Features and Generative Adversarial Network." IJACI vol.12, no.3 2021: pp.75-97. http://doi.org/10.4018/IJACI.2021070104
39. R. Ashraf, I. Kiran, T. Mahmood, A. Ur Rehman Butt, N. Razzaq and Z. Farooq, "An efficient technique for skin cancer classification using deep learning," 2020 IEEE 23rd International Multitopic Conference (INMIC), Bahawalpur, Pakistan, 2020, pp. 1-5, doi: 10.1109/INMIC50486.2020.9318164.
40. A. G. C. Pacheco and R. A. Krohling, "An Attention-Based Mechanism to Combine Images and Metadata in Deep Learning Models Applied to Skin Cancer Classification," in IEEE Journal of Biomedical and Health Informatics, vol. 25, no. 9, pp. 3554-3563, Sept. 2021, doi: 10.1109/JBHI.2021.3062002.
41. S. Alagu, K. Bhoopathy Bagan; Skin cancer classification in dermoscopy images using convolutional neural network. AIP Conference Proceedings 26 March 2021; 2336 (1): 040013. https://doi.org/10.1063/5.0045757
42. Krohling, B., Castro, P.B.C., Pacheco, A.G.C. and Krohling, R.A. (2021). A Smartphone based Application for Skin Cancer Classification Using Deep Learning with Clinical Images and Lesion Information. arXiv:2104.14353 [cs, eess]. [online] Available at: https://arxiv.org/abs/2104.14353.
43. Mijwil, M.M. Skin cancer disease images classification using deep learning solutions. Multimed Tools Appl 80, 26255–26271 (2021). https://doi.org/10.1007/s11042-021-10952-7
44. M. Shah, "LRNet: Skin Cancer Classification using Low-Resolution Images," 2021 International Conference on Communication information and Computing Technology (ICCICT), Mumbai, India, 2021, pp. 1-5, doi: 10.1109/ICCICT50803.2021.9510138.
45. Maron, R.C., Schlager, J.G., Haggenmüller, S., von Kalle, C., Utikal, J.S., Meier, F., Gellrich, F.F., Hobelsberger, S., Hauschild, A., French, L., Heinzerling, L., Schlaak, M., Ghoreschi, K., Hilke, F.J., Poch, G., Heppt, M.V., Berking, C., Haferkamp, S., Sondermann, W. and Schadendorf, D. (2021). A benchmark for neural network robustness in skin cancer classification. European Journal of Cancer, 155, pp.191–199. doi:https://doi.org/10.1016/j.ejca.2021.06.047.
46. Ali, M.S., Miah, M.S., Haque, J., Rahman, M.M. and Islam, M.K. (2021). An enhanced technique of skin cancer classification using deep convolutional neural network with transfer learning models. Machine Learning with Applications, 5, p.100036. doi:https://doi.org/10.1016/j.mlwa.2021.100036.
47. Dascalu, A., Walker, B.N., Oron, Y. et al. Non-melanoma skin cancer diagnosis: a comparison between dermoscopic and smartphone images by unified visual and sonification deep learning algorithms. J Cancer Res Clin Oncol 148, 2497–2505 (2022). https://doi.org/10.1007/s00432-021-03809-x

48. Yilmaz, A., Kalebasi, M., Samoylenko, Y., Guvenilir, M.E. and Uvet, H. (2021). Benchmarking of Lightweight Deep Learning Architectures for Skin Cancer Classification using ISIC 2017 Dataset. arXiv:2110.12270 [cs, eess]. [online] Available at: https://arxiv.org/abs/2110.12270.
49. Ahmad, B.; Jun, S.; Palade, V.; You, Q.; Mao, L.; Zhongjie, M. Improving Skin Cancer Classification Using Heavy-Tailed Student T-Distribution in Generative Adversarial Networks (TED-GAN). Diagnostics 2021, 11, 2147. https://doi.org/10.3390/diagnostics11112147
50. Kausar, N.; Hameed, A.; Sattar, M.; Ashraf, R.; Imran, A.S.; Abidin, M.Z.u.; Ali, A. Multiclass Skin Cancer Classification Using Ensemble of Fine-Tuned Deep Learning Models. Appl. Sci. 2021, 11, 10593. https://doi.org/10.3390/app112210593
51. Mazoure, B., Mazoure, A., Bédard, J. et al. DUNEScan: a web server for uncertainty estimation in skin cancer detection with deep neural networks. Sci Rep 12, 179 (2022). https://doi.org/10.1038/s41598-021-03889-2
52. Bechelli, S.; Delhommelle, J. Machine Learning and Deep Learning Algorithms for Skin Cancer Classification from Dermoscopic Images. Bioengineering 2022, 9, 97. https://doi.org/10.3390/bioengineering9030097
53. Maniraj, S.P., Maran, P.S. A hybrid deep learning approach for skin cancer diagnosis using subband fusion of 3D wavelets. J Supercomput 78, 12394–12409 (2022). https://doi.org/10.1007/s11227-022-04371-0
54. Y. Filali, H. E. Khoukhi, M. A. Sabri and A. Aarab, "Analysis and Classification of Skin Cancer Based on Deep Learning Approach," 2022 International Conference on Intelligent Systems and Computer Vision (ISCV), Fez, Morocco, 2022, pp. 1-6, doi: 10.1109/ISCV54655.2022.9806087.
55. Hassan Bedeir, R., Orban Mahmoud, R. and H. Zayed, H. (2022). Automated multi-class skin cancer classification through concatenated deep learning models. IAES International Journal of Artificial Intelligence (IJ-AI), 11(2), p.764. doi:https://doi.org/10.11591/ijai.v11.i2.pp764-772.
56. Gouda, W.; Sama, N.U.; Al-Waakid, G.; Humayun, M.; Jhanjhi, N.Z. Detection of Skin Cancer Based on Skin Lesion Images Using Deep Learning. Healthcare 2022, 10, 1183. https://doi.org/10.3390/healthcare10071183
57. Fraiwan, M.; Faouri, E. On the Automatic Detection and Classification of Skin Cancer Using Deep Transfer Learning. Sensors 2022, 22, 4963. https://doi.org/10.3390/s22134963
58. Tabrizchi, H., Parvizpour, S. & Razmara, J. An Improved VGG Model for Skin Cancer Detection. Neural Process Lett (2022). https://doi.org/10.1007/s11063-022-10927-1
59. Naeem, A.; Anees, T.; Fiza, M.; Naqvi, R.A.; Lee, S.-W. SCDNet: A Deep Learning-Based Framework for the Multiclassification of Skin Cancer Using Dermoscopy Images. Sensors 2022, 22, 5652. https://doi.org/10.3390/s22155652
60. Huynh, A.T., Hoang, VD., Vu, S., Le, T.T., Nguyen, H.D. (2022). Skin Cancer Classification Using Different Backbones of Convolutional Neural Networks. In: Fujita, H., Fournier-Viger, P., Ali, M., Wang, Y. (eds) Advances and Trends in Artificial Intelligence. Theory and Practices in Artificial Intelligence. IEA/AIE 2022. Lecture Notes in Computer Science(), vol 13343. Springer, Cham. https://doi.org/10.1007/978-3-031-08530-7_14
61. Xin, C., Liu, Z., Zhao, K., Miao, L., Ma, Y., Zhu, X., Zhou, Q., Wang, S., Li, L., Yang, F., Xu, S. and Chen, H. (2022). An improved transformer network for skin cancer classification. Computers in Biology and Medicine, 149, p.105939. doi:https://doi.org/10.1016/j.compbiomed.2022.105939.
62. Bassel, A.; Abdulkareem, A.B.; Alyasseri, Z.A.A.; Sani, N.S.; Mohammed, H.J. Automatic Malignant and Benign Skin Cancer Classification Using a Hybrid Deep Learning Approach. Diagnostics 2022, 12, 2472. https://doi.org/10.3390/diagnostics12102472
63. Ali, K., Shaikh, Z.A., Khan, A.A. and Laghari, A.A. (2022). Multiclass skin cancer classification using EfficientNets – a first step towards preventing skin cancer. Neuroscience Informatics, 2(4), p.100034. doi:https://doi.org/10.1016/j.neuri.2021.100034.
64. Qasim Gilani, S., Syed, T., Umair, M. et al. Skin Cancer Classification Using Deep Spiking Neural Network. J Digit Imaging 36, 1137–1147 (2023). https://doi.org/10.1007/s10278-023-00776-2
65. Durães, P.F.; Véstias, M.P. Smart Embedded System for Skin Cancer Classification. Future Internet 2023, 15, 52. https://doi.org/10.3390/fi15020052
66. Rezk, E., Eltorki, M. & El-Dakhakhni, W. Interpretable Skin Cancer Classification based on Incremental Domain Knowledge Learning. J Healthc Inform Res 7, 59–83 (2023). https://doi.org/10.1007/s41666-023-00127-4
67. Tembhurne, J.V., Hebbar, N., Patil, H.Y. et al. Skin cancer detection using ensemble of machine learning and deep learning techniques. Multimed Tools Appl 82, 27501–27524 (2023). https://doi.org/10.1007/s11042-023-14697-3
68. Shaaban, S., Atya, H., Mohammed, H., Sameh, A., Raafat, K., Magdy, A. (2023). Skin Cancer Detection Based on Deep Learning Methods. In: et al. The 3rd International Conference on Artificial Intelligence and Computer Vision (AICV2023), March 5–7, 2023. AICV 2023. Lecture Notes on Data Engineering and Communications Technologies, vol 164. Springer, Cham. https://doi.org/10.1007/978-3-031-27762-7_6
69. Tahir, M.; Naeem, A.; Malik, H.; Tanveer, J.; Naqvi, R.A.; Lee, S.-W. DSCC_Net: Multi-Classification Deep Learning Models for Diagnosing of Skin Cancer Using Dermoscopic Images. Cancers 2023, 15, 2179. https://doi.org/10.3390/cancers15072179
70. N. S. Karpagam, S. Sindhuja, P. Sumathi, P. Yogananth, G. Saranya and K. Madhan, "Skin Cancer Classification Based on Machine Learning Techniques," 2023 Eighth International Conference on Science Technology Engineering and Mathematics (ICONSTEM), Chennai, India, 2023, pp. 1-5, doi: 10.1109/ICONSTEM56934.2023.10142758.
71. K. Mridha, M. M. Uddin, J. Shin, S. Khadka and M. F. Mridha, "An Interpretable Skin Cancer Classification Using Optimized Convolutional Neural Network for a Smart Healthcare System," in IEEE Access, vol. 11, pp. 41003-41018, 2023, doi: 10.1109/ACCESS.2023.3269694.

72. H. L. Gururaj, N. Manju, A. Nagarjun, V. N. M. Aradhya, and F. Flammini, "DeepSkin: A Deep Learning Approach for Skin Cancer Classification," in IEEE Access, vol. 11, pp. 50205-50214, 2023, doi: 10.1109/ACCESS.2023.3274848.
73. Khan, S.A.; Gulzar, Y.; Turaev, S.; Peng, Y.S. A Modified HSIFT Descriptor for Medical Image Classification of Anatomy Objects. Symmetry 2021, 13, 1987. https://doi.org/10.3390/sym13111987
74. Gulzar, Y.; Khan, S.A. Skin Lesion Segmentation Based on Vision Transformers and Convolutional Neural Networks—A Comparative Study. Appl. Sci. 2022, 12, 5990. https://doi.org/10.3390/app12125990
75. Khan, F.; Ayoub, S.; Gulzar, Y.; Majid, M.; Reegu, F.A.; Mir, M.S.; Soomro, A.B.; Elwasila, O. MRI-Based Effective Ensemble Frameworks for Predicting Human Brain Tumor. J. Imaging 2023, 9, 163. https://doi.org/10.3390/jimaging9080163
76. Mehmood, A.; Gulzar, Y.; Ilyas, Q.M.; Jabbari, A.; Ahmad, M.; Iqbal, S. SBXception: A Shallower and Broader Xception Architecture for Efficient Classification of Skin Lesions. Cancers 2023, 15, 3604. https://doi.org/10.3390/cancers15143604
77. N. Siddique, S. Paheding, C. P. Elkin and V. Devabhaktuni, "U-Net and Its Variants for Medical Image Segmentation: A Review of Theory and Applications," in IEEE Access, vol. 9, pp. 82031-82057, 2021, doi: 10.1109/ACCESS.2021.3086020.
78. Krithika alias AnbuDevi, M.; Suganthi, K. Review of Semantic Segmentation of Medical Images Using Modified Architectures of UNET. Diagnostics 2022, 12, 3064. https://doi.org/10.3390/diagnostics12123064
79. Sreelatha, T., Subramanyam, M.V. & Prasad, M.N.G. Early Detection of Skin Cancer Using Melanoma Segmentation technique. J Med Syst 43, 190 (2019). https://doi.org/10.1007/s10916-019-1334-1
80. L. Liu, L. Mou, X. X. Zhu and M. Mandal, "Skin Lesion Segmentation Based on Improved U-net," 2019 IEEE Canadian Conference of Electrical and Computer Engineering (CCECE), Edmonton, AB, Canada, 2019, pp. 1-4, doi: 10.1109/CCECE.2019.8861848.
81. J. Wu, E. Z. Chen, R. Rong, X. Li, D. Xu and H. Jiang, "Skin Lesion Segmentation with C-UNet," 2019 41st Annual International Conference of the IEEE Engineering in Medicine and Biology Society (EMBC), Berlin, Germany, 2019, pp. 2785-2788, doi: 10.1109/EMBC.2019.8857773.
82. Tang, P., Liang, Q., Yan, X., Xiang, S., Sun, W., Zhang, D., Coppola, G., 2019. Efficient skin lesion segmentation using separable-Unet with stochastic weight averaging. Computer Methods and Programs in Biomedicine 178, 289–301. https://doi.org/10.1016/j.cmpb.2019.07.005
83. Araújo, R.L., Araújo, F.H.D.d. & Silva, R.R.V.e. Automatic segmentation of melanoma skin cancer using transfer learning and fine-tuning. Multimedia Systems 28, 1239–1250 (2022). https://doi.org/10.1007/s00530-021-00840-3
84. Nawaz, M., Nazir, T., Masood, M., Ali, F., Khan, M.A., Tariq, U., Sahar, N., Damaševičius, R., 2022. Melanoma segmentation: A framework of improved DenseNet77 and UNET convolutional neural network. International Journal of Imaging Systems and Technology 32, 2137–2153. https://doi.org/10.1002/ima.22750
85. Chincholi, F.; Koestler, H. Detectron2 for Lesion Detection in Diabetic Retinopathy. Algorithms 2023, 16, 147. https://doi.org/10.3390/a16030147